\documentclass{article}

 \usepackage[preprint]{neurips_2026}

\usepackage[utf8]{inputenc} % allow utf-8 input
\usepackage[T1]{fontenc}    % use 8-bit T1 fonts
\usepackage{hyperref}       % hyperlinks
\usepackage{url}            % simple URL typesetting
\usepackage{booktabs}       % professional-quality tables
\usepackage{amsfonts}       % blackboard math symbols
\usepackage{nicefrac}       % compact symbols for 1/2, etc.
\usepackage{microtype}      % microtypography
\usepackage{xcolor}         % colors
\usepackage{amsmath}

\usepackage{graphicx}
\usepackage{listings}
\usepackage{multirow}
\usepackage{makecell}
\usepackage{subcaption}
\usepackage{enumitem}
\usepackage{tikz}
\usepackage{cleveref}
\usepackage{debate}

\usetikzlibrary{shapes, arrows.meta, positioning, fit, backgrounds}

\title{DetectZoo: A Unified Toolkit for AI-Generated Content Detection Across Text, Audio, and Image Modalities}

\author{%
Sajad Ebrahimi$^1$ \quad
Nima Jamali$^2$ \quad
Bardia Shirsalimian$^3$ \quad
Kelly McConvey$^1$ \quad \\
\textbf{Wentao Zhang}$^2$ \quad
\textbf{Jalehsadat Mahdavimoghaddam}$^1$ \quad
\textbf{Maksym Taranukhin}$^{4,5}$ \quad \\
\textbf{Maura Grossman}$^2$ \quad 
\textbf{Vered Shwartz}$^{4,5}$ \quad
\textbf{Yuntian Deng}$^2$ \quad
\textbf{Ebrahim Bagheri}$^1$ \\
$^1$University of Toronto \quad
$^2$University of Waterloo \quad
$^3$Toronto Metropolitan University \quad \\
$^4$University of British Columbia  \quad
$^5$Vector Institute \\ 
\texttt{\{s.ebrahimi, kelly.mcconvey, jaleh.mahdavimoghaddam, ebrahim.bagheri\}@utoronto.ca} \\
\texttt{\{nima.jamali, w564zhan, maura.grossman, yuntian\}@uwaterloo.ca}\\
\texttt{bardia.shirsalimian@torontomu.ca}\\
\texttt{\{maksymt, vshwartz\}@cs.ubc.ca}\\
}

\begin{document}

\maketitle

\begin{abstract}
The growing popularity and capacity of generative models have eroded the distinction between human and machine-generated content, motivating a growing body of work on detection across text, images, and audio. Most available detectors are either commercial software or, if open-source, come with incompatible codebases with bespoke preprocessing, evaluation protocols, and evaluation metrics, which make their adoption, fair comparison, and reproduction quite difficult. To address this critical gap, we introduce \texttt{DetectZoo}, a first-of-its-kind, extensible toolkit designed to provide a unified interface for AI-generated content detection across text, audio, and image modalities. \texttt{DetectZoo} standardizes the complete empirical pipeline, from data ingestion and preprocessing to model assessment, offering researchers a cohesive framework to benchmark state-of-the-art detectors systematically. By integrating diverse public datasets and baseline detection algorithms under a single, unified API, our toolkit facilitates rigorous and reproducible evaluation. \texttt{DetectZoo} provides reference implementations of \textit{61 detectors}, native loaders for \textit{22 benchmark datasets}, and a standardized evaluation pipeline that reports multiple metrics through a common interface. Each detector is self-contained yet accessible through the same interface, automatically caches pretrained weights, and reproduces the original published results. \texttt{DetectZoo} lowers the barrier to entry for multi-modal AI forensics, enabling researchers to identify performance gaps across domains and accelerating the development of robust, generalizable detection techniques. The open-source repository and comprehensive documentation are publicly available at \url{https://github.com/sadjadeb/DetectZoo}, and the package can be installed via \href{https://pypi.org/project/detectzoo/}{\texttt{pip install detectzoo}}.
\end{abstract}

\section{Introduction}
 
The rapid advancement of generative AI has made the creation of highly realistic synthetic content trivially accessible. Modern architectures, including Large Language Models, latent diffusion models, and neural audio codecs, can synthesize highly realistic text, images, and speech at unprecedented scales. Malicious actors increasingly leverage synthetic content to fabricate misleading news articles~\cite{zellers2019defending}, generate non-consensual deepfake imagery~\cite{Croitoru2024DeepfakeMG, chesney2019deep}, synthesize spoofed voice recordings for fraud~\cite{yi2023audio, muller2022human}, and manufacture fabricated evidence for use in judicial proceedings~\cite{apolo2024beyond, grossman2024judicial}. Consequently, distinguishing human from machine-generated content has become a core problem in AI safety.
% and digital forensics.
 
A growing body of research has responded by developing detection methods for AI-generated content across text~\cite{mitchell2023detectgpt,bao2023fast,hans2024binoculars}, images~\cite{ojha2023towards, chen2024drct, cheng2025co}, and audio~\cite{tak2021rawnet2,jung2022aasist}. Each new method is typically released as a standalone codebase with its own preprocessing pipeline, evaluation datasets, and metrics implementations. This fragmentation produces three obstacles for the community to collectively move forward: (1) \textit{Lack of fair comparison.} Differences in tokenization, image preprocessing, audio resampling, threshold selection, and train and test splits make reported numbers across papers incommensurable. (2) \textit{Reproducibility barriers.} Missing dependencies, undocumented hyperparameters, and particular code structure mean that reproducing even a single method's published results often requires significant engineering effort. (3) \textit{No cross-modal perspective.} Detectors for text, images, and audio exist in entirely separate ecosystems, which prevents the unified study of AI-generated content detection as a coherent research problem.

\begin{figure}
    \centering
    \includegraphics[width=\linewidth]{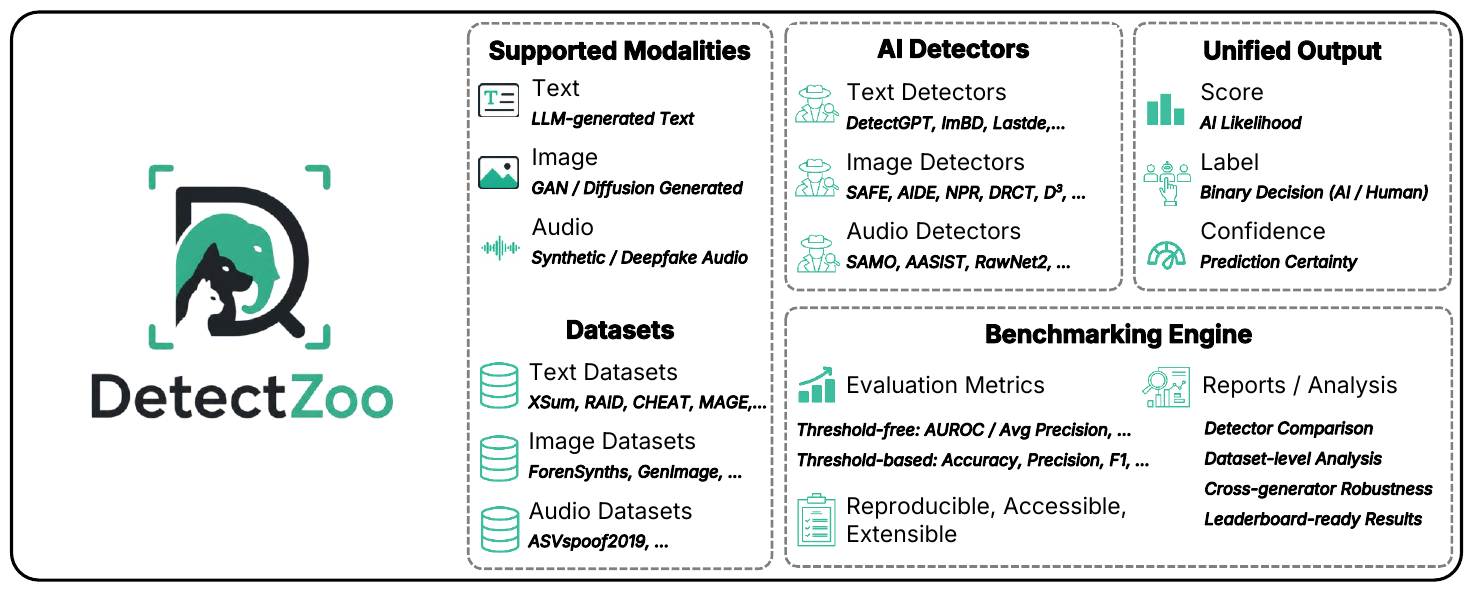}
    \caption{Overview of \texttt{DetectZoo}. The framework standardizes the evaluation of AI-generated content detectors across text, image, and audio. It provides a unified API from data ingestion to metric computation.}
    \label{fig:overview}
\end{figure}
 
We address these obstacles with \texttt{DetectZoo}, a unified, multi-modal toolkit for the evaluation of AI-generated content detectors. \texttt{DetectZoo} aggregates \textit{61 detection methods} across \textit{three modalities} into a single codebase with a consistent interface, paired with \textit{22 benchmark datasets} and a standardized evaluation pipeline. By seamlessly integrating different modalities into a single extensible repository, \texttt{DetectZoo} empowers researchers to conduct rigorous comparisons of state-of-the-art detection algorithms under standardized experimental conditions. \Cref{fig:overview} provides an overview of the \texttt{DetectZoo} framework. Specifically, our primary contributions are as follows:

\begin{itemize}
    \item \textbf{Unified Multi-Modal API:} We provide a standardized architecture that spans the text, image, and audio detection methods. This abstraction significantly reduces the engineering overhead required to benchmark novel detection algorithms against established baselines.
    \item \textbf{Standardized Benchmark Datasets:} We curate and natively integrate a diverse suite of challenging public datasets spanning all three modalities, ensuring robust and reproducible evaluation environments.
    \item \textbf{Comprehensive Baseline Implementations:} We implement a wide array of state-of-the-art detection baselines, covering approaches from zero-shot statistical anomaly detection to supervised deep representation learning, facilitating immediate comparative analysis.
    \item \textbf{Open-Source Extensibility:} We release the complete toolkit, including modular evaluation pipelines and extensive documentation, to foster community-driven advancements and accelerate the development of generalizable AI forensics.
\end{itemize}

We should note that \texttt{DetectZoo} is not a new detection algorithm; it is a research infrastructure contribution. Its value lies in making detection methods \emph{accessible}, \emph{comparable}, and \emph{reproducible}, in the same spirit as CLIMB~\cite{liu2025climb} for class-imbalanced learning, RobustBench~\cite{croce2021robustbench} for adversarial robustness, Hugging Face Transformers~\cite{wolf2019huggingface} for NLP models, and PyOD~\cite{zhao2019pyod} for outlier detection. We believe that a community-maintained toolkit advances AI-generated content detection more effectively than isolated per-paper codebases. 

% =====================================================================
% 2. Related Work
% =====================================================================
\section{Related Work}

The challenge of detecting AI-generated content has encouraged significant research efforts across multiple independent domains. We categorize the existing literature based on the primary modality of focus, highlighting the inherent fragmentation that \texttt{DetectZoo} seeks to resolve.
 
\subsection{Detectors}
 
\paragraph{Text Detectors.} Text detection methods fall into five paradigms.
\emph{Zero-shot statistical methods} operate on the intuition that machine-generated text favors predictable probability distributions and highly likely word choices compared to the natural variance of human writing. To capture this, these methods evaluate token-level signals, such as log-likelihood~\cite{solaiman2019release}, rank~\cite{gehrmann2019gltr}, entropy~\cite{lavergne2008entropy}, or their respective ratios (e.g., LRR~\cite{su2023detectllm}), utilizing a reference language model.
\emph{Perturbation-based methods} are grounded in the observation that AI-generated text often occupies local maxima within a model's probability landscape, meaning that minor alterations typically degrade the sequence likelihood more sharply than they would for human text. Accordingly, these methods quantify probability curvature by contrasting original passages with altered variants. For instance, DetectGPT~\cite{mitchell2023detectgpt} employs T5 for perturbation generation, whereas Fast-DetectGPT~\cite{bao2023fast} analytically approximates this curvature to bypass explicit text alteration.
\emph{Multi-model techniques} rely on the premise that different language models share underlying structural biases, which can be exposed by cross-referencing text against diverse models to find common artificial patterns. These techniques exploit multiple generative architectures or iterative rewriting operations. Prominent examples include Binoculars~\cite{hans2024binoculars}, which contrasts the perplexities of two distinct models, and DNA-GPT~\cite{yang2023dna}, which analyzes the divergence of regenerated continuations.
\emph{Representation-based approaches} investigate the geometric properties of hidden states; for example, PHD~\cite{tulchinskii2023intrinsic} estimates the intrinsic dimensionality of token embeddings via persistent homology. 
Finally, \emph{supervised methods} take the direct approach of explicitly teaching a neural network to recognize the stylistic and structural hallmarks of synthetic text. These methods rely on fine-tuning discriminative classifiers over labeled corpora, encompassing models like RoBERTa-based detectors~\cite{solaiman2019release}, adversarial frameworks such as RADAR~\cite{hu2023radar}, and alignment-focused reward models~\cite{lee2024remodetect}.
 
\paragraph{Image Detectors.}
AI-generated image detectors target artifacts introduced by generative pipelines. 
\emph{Artifact-based methods} learn discriminative pixel-space inconsistencies. CNNSpot~\cite{wang2020cnn} fine-tunes a ResNet-50 with strong augmentations, while NPR~\cite{tan2024rethinking} exposes upsampling artifacts via nearest-neighbor down-up residuals.
\emph{Frequency- and structure-aware methods} detect synthetic artifacts that manifest as abnormal spectral or derivative-domain patterns. FreqNet \cite{tan2024frequency} employs frequency-domain representations to capture generator-specific spectral artifacts, SAFE~\cite{li2025improving} isolates high-frequency components using Discrete Wavelet Transforms (DWT), while LGrad~\cite{tan2023learning} leverages gradient-domain representations to highlight structural inconsistencies.
\emph{Reconstruction-based methods} measure discrepancies between an image and its re-encoded counterpart, where larger reconstruction errors indicate weaker compatibility with the generative model. For instance, AEROBLADE~\cite{ricker2024aeroblade} computes Learned Perceptual Image Patch Similarity (LPIPS) \cite{zhang2018unreasonable} distance across a Variational Autoencoder (VAE) round-trip.
\emph{CLIP-based methods} leverage pretrained vision-language representations for real-vs-fake classification. UnivFD~\cite{ojha2023towards} trains a linear classifier on top of frozen CLIP embeddings, FatFormer~\cite{liu2024forgery} augments CLIP with frequency-aware adapters, and C2P-CLIP~\cite{tan2025c2p} adapts CLIP via prompt learning for improved real vs. fake discrimination.
\emph{Contrastive methods} such as DRCT~\cite{chen2024drct} learn transferable artifact representations using diffusion-reconstruction pairs, while \emph{hybrid methods} like AIDE~\cite{yansanity} combine SRM-based frequency cues with large-scale semantic backbones.
 
\paragraph{Audio Detectors.}  Audio deepfake detection primarily follows two overarching paradigms.
\emph{End-to-end waveform methods} operate on the intuition that artificial synthesis engines introduce microscopic, sample-level acoustic anomalies that are frequently discarded during traditional feature extraction. Consequently, these approaches operate directly on the raw acoustic signal. For instance, RawNet2~\cite{tak2021rawnet2} uses sinc-filter front-ends, and AASIST~\cite{jung2022aasist} adds spectro-temporal graph attention.
\emph{Pretrained feature methods} rely on the premise that large-scale foundation models capture rich representations of natural human speech, making their latent embeddings highly sensitive to the subtle phonetic and structural deviations characteristic of synthetic voices. These approaches utilize representations extracted from foundation audio models. As an illustration, Whisper-MesoNet~\cite{kawa23b_interspeech} uses Whisper encoder embeddings.

\subsection{Benchmarks and Evaluation Toolkits}
\label{sec:related:benchmarks}

A range of benchmarks and toolkits within the literature have evaluated AI-generated content detection, though typically within isolated modalities. In the text domain, resources like RAID~\cite{dugan2024raid} and TuringBench~\cite{uchendu2021turingbench} focus heavily on massive data curation. RAID provides over ten million documents spanning eleven language models and twelve adversarial attacks, while TuringBench evaluates outputs from nineteen distinct neural generators. Among text-based benchmarks, RAID distinguishes itself as the only benchmark that provides a dedicated Python package for an end-to-end, standardized evaluation pipeline. M4~\cite{wang2024m4} further expands evaluation to multi-domain and multilingual environments. Toolkits such as MGTBench~\cite{he2024mgt} build upon these data efforts by offering integrated scripts to evaluate 14 distinct detectors. For visual media, DeepfakeBench~\cite{yan2023deepfakebench} provides a comprehensive suite by unifying 28 image detectors across 8 datasets. Similarly, widely used multi-generator testbeds like GenImage~\cite{zhu2023genimage}, ForenSynths~\cite{wang2020cnn}, and AIGCDetection~\cite{zhong2023patchcraft} supply evaluation environments that are sometimes coupled with built-in baseline algorithms. In audio forensics, the ASVspoof challenge~\cite{todisco2019asvspoof} and its corresponding baselines~\cite{yamagishi21_asvspoof} establish the standard benchmarking environment by evaluating core detection methods against known spoofing techniques.

Although these frameworks provide valuable curated datasets and baseline algorithms, they remain strictly fragmented by modality. Many data-focused benchmarks distribute labeled artifacts but omit the corresponding detection algorithms, requiring researchers to independently implement and test methods on their own. Conversely, existing toolkits often lack standardized programming interfaces or automated data retrieval. \texttt{DetectZoo} bridges this gap by introducing a comprehensive cross-modal toolkit that seamlessly integrates these datasets via built-in loaders. It unifies 61 baseline detectors and 22 datasets across text, images, and audio into a single, cohesive API with fully automated data management. Table~\ref{tab:toolkit_comparison} summarizes the comparison.

\begin{table}[t]
\centering
\caption{Comparison of \texttt{DetectZoo} with existing forensic detection toolkits. ``Unified API'' indicates a single load and predict interface across detectors. ``Auto-DL'' indicates that benchmark datasets download and cache automatically. Detector and Dataset counts reflect the most recent publicly tagged release of each toolkit.}
\label{tab:toolkit_comparison}
\resizebox{\textwidth}{!}{
\begin{tabular}{lccccccc}
\toprule
Benchmark/Toolkit                                                   & Text                                          & Image                     & Audio                     & \# Detectors & \# Datasets & Unified API               & Auto-DL                   \\ 
\midrule
TuringBench~\cite{uchendu2021turingbench}       & \checkmark                     &                           &                           & 10           & 1           &                           &  \\
MGTBench~\cite{he2024mgt}                     & \checkmark                     &                           &                           & 14           & 3           &  \checkmark                         & \checkmark \\
M4~\cite{wang2024m4}                     & \checkmark                     &                           &                           & 6           & 1           &                           & \checkmark \\
MAGE~\cite{li2024mage}                     & \checkmark                     &                           &                           & 4           & 10           &                           & \checkmark \\
RAID~\cite{dugan2024raid}                     & \checkmark                     &                           &                           & 14           & 1           &   \checkmark                        & \checkmark \\
DetectRL~\cite{wu2024detectrl}                & \checkmark &       &      & 15           & 4           &       
&    \\
DeepfakeBench~\cite{yan2023deepfakebench}                        &                                               & \checkmark                           &                           & 28             & 8             & \checkmark        
&  \\
GenImage~\cite{zhu2023genimage}                             &                                               & \checkmark                          &                           & 7             & 1            &           
&  \\
AIGCDetection~\cite{zhong2023patchcraft}                     &                                               & \checkmark                          &                           & 10             & 1            &              
&   \\
AIGIBench~\cite{liartificial}                     &                                               & \checkmark                          &                           & 12             & 1            &                           &                           \\
ASVspoof Baselines~\cite{yamagishi21_asvspoof} &                &                           &  \checkmark                         & 4             & 1            &                           &                           \\ 
\midrule
\textbf{DetectZoo (ours)}                                           & \checkmark                     & \checkmark & \checkmark & \textbf{61}  & \textbf{22} & \checkmark & \checkmark \\ 
\bottomrule
\end{tabular}
}
\end{table}

% =====================================================================
% 3. The DetectZoo Toolkit
% =====================================================================
\section{The \texttt{DetectZoo} Toolkit}

To overcome the systemic fragmentation detailed in the preceding sections, we introduce \texttt{DetectZoo}. Engineered with a rigorous focus on utility and reproducibility, this toolkit provides a unified and extensible infrastructure for benchmarking AI-generated content detectors across text, image, and audio modalities. It is designed around three core principles of reproducibility, accessibility, and extensibility.

% which are realized through the modular architecture shown in Figure~\ref{fig:architecture}. 
\textit{Reproducibility} is supported by providing self-contained implementations of established detectors, addressing common challenges such as missing implementation details and unstable dependencies. The toolkit automatically manages pretrained weights and model components to ensure a stable evaluation environment, and each detector is validated against reported results from original works within a documented tolerance as described in Section~\ref{sec:reproduction}. 
\textit{Accessibility} is achieved through a unified interface where all detectors are accessed through a single factory function and operate on diverse input modalities while returning standardized prediction objects. This abstraction enables consistent interaction across heterogeneous models without requiring users to manage modality-specific differences.
\textit{Extensibility} is the central design goal that allows researchers to seamlessly incorporate new detectors and datasets. The framework adopts a lightweight registration-based pattern in which new components are introduced by subclassing core abstractions such as \texttt{BaseDetector} or modality-specific variants such as \texttt{BaseTextDetector} and annotating them with decorators like \texttt{@register\_detector}. Once registered, these components are immediately accessible through the same factory interface, ensuring uniform integration. The same pattern applies to datasets via \texttt{@register\_dataset}. To support practical adoption, \texttt{DetectZoo} is distributed as a PyPI package \texttt{detectzoo} with modular optional dependencies, enabling lightweight installations tailored to specific use cases. Furthermore, the repository includes dedicated modules that leverage \texttt{DetectZoo}'s unified functions to directly reproduce the empirical results originally reported in previous benchmarking papers.

\subsection{The Unified Application Programming Interface}
 
The primary technical innovation of \texttt{DetectZoo} is its Unified Application Programming Interface. By standardizing the experimental lifecycle, the API dramatically reduces the friction traditionally associated with cross-modal benchmarking. All detectors adhere to a strictly typed \texttt{predict()} method that returns a standardized \texttt{DetectionResult} dataclass. This object encapsulates the binary prediction label, a continuous anomaly score, prediction confidence, and a flexible metadata dictionary for algorithm-specific metrics. Consequently, researchers can load varied models, from zero-shot statistical tools to supervised neural networks, using a single \texttt{load\_detector} functional call. 
The provided code snippet in~\Cref{lst:quickstart} demonstrates this simplicity.

\begin{lstlisting}[caption={One-line loading and uniform inference across modalities. The same \texttt{predict()} call accepts a string, an image path, or an audio path.}, label=lst:quickstart, float=t]
from detectzoo import load_detector
 
text_det  = load_detector("fast_detectgpt", device="cuda")
image_det = load_detector("aeroblade")
audio_det = load_detector("rawnet2")
 
text_result  = text_det.predict("This passage was written by an LLM.")
image_result = image_det.predict("path/to/image.png")
audio_result = audio_det.predict("path/to/clip.wav")
 
print(text_result)
# DetectionResult(score=1.2345, label='ai', confidence=0.8012)
\end{lstlisting}

\subsection{Evaluation Pipeline}
\label{sec:design:eval}
 
\paragraph{Benchmark Evaluator.}
The core of the empirical benchmarking process is managed by the \texttt{BenchmarkEvaluator} module. This component accepts a specified dataset alongside a suite of initialized detectors. It systematically iterates through the data, leveraging the unified \texttt{predict()} interface to aggregate continuous anomaly scores and ground-truth labels, which are subsequently used to compute a comprehensive array of performance metrics.

\begin{comment}
\begin{lstlisting}[caption={Running a multi-detector benchmark. The same \texttt{BenchmarkEvaluator} handles any modality.}, label=lst:benchmark, float=t]
from detectzoo import load_detector, load_dataset
from detectzoo.benchmarks import BenchmarkEvaluator
 
dataset = load_dataset("hc3", max_samples=1000)
 
detectors = [
    load_detector("binoculars",     device="cuda"),
    load_detector("fast_detectgpt", device="cuda"),
    load_detector("radar",          device="cuda"),
]
 
evaluator = BenchmarkEvaluator(dataset)
results = evaluator.run_and_save(detectors, output_path="results/hc3.json")
\end{lstlisting}
 \end{comment}
 
\paragraph{Metrics.}
The pipeline computes both threshold-independent and threshold-dependent statistics. Threshold-independent measures include the Area Under the Receiver Operating Characteristic curve (AUROC), Precision-Recall AUC, Average Precision, and the Equal Error Rate (EER). The EER is linearly interpolated along the ROC curve to match the convention used in audio anti-spoofing research. Threshold-dependent measures include accuracy, precision, recall, the $F_1$ score, True Positive Rate, and False Positive Rate.
 
\paragraph{Experimental Logging.}
\texttt{DetectZoo} exports every evaluation outcome to a structured JSON file that records the computed metrics together with experimental metadata (dataset specification, detector configurations, sample counts, and hyperparameters), so that benchmarking experiments are self-documenting and replayable.

\subsection{Implemented Detection Methods}
\label{sec:methods}
 \texttt{DetectZoo} aggregates 61 detection methods spanning text, image, and audio modalities, with each algorithm implemented as a self-contained unit that vendors its required components and caches pretrained weights upon first use. The text modality includes 36 detectors covering a broad spectrum of paradigms, ranging from zero-shot statistical estimators to supervised models. These detectors operate directly on raw string inputs, with the framework abstracting tokenizer initialization, sequence truncation, and context window management. The implemented text detectors are summarized in Table~\ref{tab:text_detectors}, and their outputs are validated against the original authors' reported results as discussed in Section~\ref{sec:reproduction}.

\begin{table}[t]
\centering
\caption{\textbf{Text detectors in \texttt{DetectZoo}} (36 methods). Category abbreviations: \textbf{Stat} = Zero-shot Statistical; \textbf{Pert} = Perturbation/Distribution-based; \textbf{Multi} = Multi-model/Generation-based; \textbf{Repr} = Layer/Representation Analysis; \textbf{Sup} = Supervised/Reward-model; \textbf{OOD} = Out-of-Distribution.}
\label{tab:text_detectors}
\resizebox{\textwidth}{!}{%
\begin{tabular}{@{}llll@{}}
\toprule
\textbf{Category} & \textbf{Registry Name} & \textbf{Method Explanation} & \textbf{Reference} \\
\midrule
\multirow{8}{*}{Stat}
 & \texttt{log\_likelihood} & Average token log-probability & Gehrmann et al.~\cite{gehrmann2019gltr} \\
 & \texttt{log\_rank}       & Average log-rank of observed tokens & Gehrmann et al.~\cite{gehrmann2019gltr} \\
 & \texttt{rank}            & Average raw token dictionary rank & Gehrmann et al.~\cite{gehrmann2019gltr} \\
 & \texttt{entropy}         & Average predictive token entropy & Gehrmann et al.~\cite{gehrmann2019gltr} \\
 & \texttt{lrr}             & Log-Likelihood Ratio normalized by rank & Su et al.~\cite{su2023detectllm} \\
 & \texttt{lastde}          & Multiscale diversity entropy of token sequences & Xu et al.~\cite{xu2024lastde} \\
 & \texttt{gecscore}        & Grammar correction frequency & Wu et al.~\cite{wu2025gecscore} \\
 & \texttt{biscope}         & Bidirectional cross-entropy & Guo et al.~\cite{guo2024biscope} \\
\midrule
\multirow{6}{*}{Pert}
 & \texttt{detectgpt}       & Log-probability curvature via perturbations & Mitchell et al.~\cite{mitchell2023detectgpt} \\
 & \texttt{fast\_detectgpt} & Perturbation-free curvature estimation & Bao et al.~\cite{bao2023fast} \\
 & \texttt{adadetectgpt}    & Adaptive curvature using B-spline witness & Zhou et al.~\cite{zhou2025adadetect} \\
 & \texttt{npr}             & Normalized Perturbation Rank across mutations & Su et al.~\cite{su2023detectllm} \\
 & \texttt{lastde\_pp}      & Distributional entropy on perturbed samples & Xu et al.~\cite{xu2024lastde} \\
 & \texttt{glimpse}         & Probability distribution estimation and curvature & Bao et al.~\cite{bao2025glimpse} \\
\midrule
\multirow{8}{*}{Multi}
 & \texttt{binoculars}      & Ratio of log-likelihoods between two LLMs & Hans et al.~\cite{hans2024binoculars} \\
 & \texttt{dna\_gpt}        & Divergent n-gram analysis of LLM continuations & Yang et al.~\cite{yang2023dna} \\
 & \texttt{dna\_detectllm}  & Mutation-repair paradigm for sequence identity & Zhu et al.~\cite{zhu2025dnadetect} \\
 & \texttt{revise\_detect}  & Semantic invariance under LLM revision & Zhu et al.~\cite{zhu2023revise} \\
 & \texttt{raidar}          & Rewriting-invariance via alignment scoring & Mao et al.~\cite{mao2024raidar} \\
 & \texttt{ghostbuster}     & Features from multiple weak language models & Verma et al.~\cite{verma2024ghostbuster} \\
 & \texttt{tocsin}          & Token cohesiveness via BARTScore & Ma \& Wang~\cite{ma-wang2024tocsin} \\
 & \texttt{ipad}            & Consistency check via inverse prompting & Chen et al.~\cite{chen2025ipad} \\
\midrule
\multirow{4}{*}{Repr}
 & \texttt{text\_fluoroscopy} & Layer-wise KL divergence analysis & Yu et al.~\cite{yu2024text} \\
 & \texttt{coco}             & Contrastive inter-sentence coherence & Liu et al.~\cite{liu2023coco} \\
 & \texttt{phd}              & Persistent Homology of embedding dimensions & Tulchinskii et al.~\cite{tulchinskii2023intrinsic} \\
 & \texttt{mle\_ide}         & MLE-based intrinsic dimension estimation & Tulchinskii et al.~\cite{tulchinskii2023intrinsic} \\
\midrule
\multirow{7}{*}{Sup}
 & \texttt{roberta\_base}   & Fine-tuned RoBERTa-base classifier by OpenAI & Solaiman et al.~\cite{solaiman2019release} \\
 & \texttt{roberta\_large}  & Fine-tuned RoBERTa-large classifier by OpenAI & Solaiman et al.~\cite{solaiman2019release} \\
 & \texttt{radar}           & Adversarial paraphrase-robust discriminator & Hu et al.~\cite{hu2023radar} \\
 & \texttt{imbd}            & Semantic noise filtering via imitation modeling & Chen et al.~\cite{chen2025imitate} \\
 & \texttt{remodetect}      & Reward-model scoring (DeBERTa-v3) & Lee et al.~\cite{lee2024remodetect} \\
 & \texttt{detective}       & Multi-level contrastive learning with KNN & Guo et al.~\cite{guo2024detective} \\
 & \texttt{irm}             & Implicit reward modeling via DPO ratios & Liu et al.~\cite{liu2026irm} \\
\midrule
\multirow{3}{*}{OOD}
 & \texttt{dsvdd}           & One-class hypersphere anomaly detection & Zeng et al.~\cite{zeng2025ood} \\
 & \texttt{hrn}             & Holistic Regularized Network for OOD & Zeng et al.~\cite{zeng2025ood} \\
 & \texttt{energy\_detector}& Scalar energy-based OOD scoring & Zeng et al.~\cite{zeng2025ood} \\
\bottomrule
\end{tabular}%
}
\end{table}

The image modality comprises 15 detectors that capture diverse methodological approaches such as frequency domain spectral analysis, latent reconstruction error modeling, vision language probing, and diffusion based contrastive learning. The vision interface accepts both raw file paths and PIL image objects, and enforces consistent preprocessing including spatial normalization, tensor formatting, and backend specific transformations. The implemented image detectors are summarized in Table~\ref{tab:image_detectors}.

\begin{table}[t]
\centering
\caption{\textbf{Image detectors in \texttt{DetectZoo}} (15 methods).}
\label{tab:image_detectors}
\resizebox{\textwidth}{!}{%
\begin{tabular}{@{}lll@{}}
\toprule
\textbf{Registry Name} & \textbf{Method Explanation} & \textbf{Reference} \\
\midrule
\texttt{aeroblade}      & VAE reconstruction LPIPS distance & Ricker et al.~\cite{ricker2024aeroblade} \\
\texttt{aide}           & Hybrid frequency-patch and CLIP features &  Yan et al.~\cite{yansanity} \\
\texttt{cnnspot}        & ResNet-50 trained on ProGAN \cite{karras2018progressive} & Wang et al.~\cite{wang2020cnn} \\
\texttt{c2p\_clip}      & CLIP with category-common injection & Tan et al.~\cite{tan2025c2p} \\
\texttt{cospy}          & Adaptive fusion of semantic and pixel artifacts & Cheng et al.~\cite{cheng2025co} \\
\texttt{d3}             & Dual-branch discrepancy learning from distortions & Yang et al.~\cite{yang2025d} \\
\texttt{drct}           & Diffusion Reconstruction Contrastive Training & Chen et al.~\cite{chen2024drct} \\
\texttt{fatformer}      & Forgery-aware CLIP adaptation with image/frequency cues & Liu et al.~\cite{liu2024forgery} \\
\texttt{freqnet}        & CNN-based frequency-domain detection & Tan et al.~\cite{tan2024frequency} \\
\texttt{lgrad}          & Gradient-space synthetic artifact detection & Tan et al.~\cite{tan2023learning} \\
\texttt{manifold\_bias} & Score-function manifold bias analysis & Brokman et al.~\cite{brokman2025manifold} \\
\texttt{npr\_deepfake}  & Neighboring pixel relationship artifacts & Tan et al.~\cite{tan2024rethinking} \\
\texttt{patchcraft}     & Contrastive poor/rich texture patch analysis & Zhong et al.~\cite{zhong2023patchcraft} \\
\texttt{safe}           & Bias-reduced local artifact learning & Li et al.~\cite{li2025improving} \\
\texttt{univfd}         & Linear probe on frozen CLIP-ViT features & Ojha et al.~\cite{ojha2023towards} \\
\bottomrule
\end{tabular}%
}
\end{table}

The audio modality includes 10 architectures designed for detecting synthetic speech and deepfake voice signals, where the ingestion pipeline supports both file-based inputs and waveform representations paired with sampling rates, while internally handling resampling and temporal alignment through integrations with libraries such as \texttt{torchaudio} and \texttt{librosa}. The implemented audio detectors are summarized in Table~\ref{tab:audio_detectors}.

\begin{table}[t]
\centering
\caption{\textbf{Audio detectors in \texttt{DetectZoo}} (10 methods).}
\label{tab:audio_detectors}
\resizebox{\textwidth}{!}{
\begin{tabular}{@{}lll@{}}
\toprule
\textbf{Registry Name} & \textbf{Method Explanation} & \textbf{Reference} \\
\midrule
\texttt{rawnet2}
  & End-to-end sinc-filter front-end with residual blocks and GRU
  & Tak et al.~\cite{tak2021rawnet2} \\
\texttt{aasist}
  & Integrated spectro-temporal graph attention network
  & Jung et al.~\cite{jung2022aasist} \\
\texttt{rawgat\_st}
  & End-to-end spectro-temporal graph attention on raw waveform
  & Tak et al.~\cite{tak2021rawgat} \\
\texttt{res\_tssdnet}
  & Residual time-domain and spectral-domain dilated network
  & Hua et al.~\cite{hua2021tssdnet} \\
\texttt{samo}
  & Speaker attractor multi-center one-class learning
  & Ding et al.~\cite{ding2023samo} \\
\texttt{ast\_asvspoof}
  & Audio Spectrogram Transformer fine-tuned on ASVspoof 2019
  & Gong et al.~\cite{gong2021ast} \\
\midrule
\texttt{anti\_deepfake\_wav2vec}
  & SSL post-training of Wav2Vec2-Large on 74k hrs speech
  & Ge et al.~\cite{ge2025antideepfake} \\
\texttt{anti\_deepfake\_hubert}
  & SSL post-training of HuBERT-XLarge on 74k hrs speech
  & Ge et al.~\cite{ge2025antideepfake} \\
\texttt{anti\_deepfake\_xlsr2b}
  & SSL post-training of XLS-R-2B on 74k hrs speech
  & Ge et al.~\cite{ge2025antideepfake} \\
\texttt{xlsr\_sls}
  & Sensitive layer selection over XLS-R backbone
  & Zhang et al.~\cite{zhang2024xlsrsls} \\
\bottomrule
\end{tabular}
}
\end{table}

In addition to detection methods, \texttt{DetectZoo} provides integrated support for 22 benchmark datasets across all three modalities as listed in Table~\ref{tab:datasets}. Each dataset is automatically downloaded upon first use and cached in a configurable directory, eliminating the need for manual data preparation. All datasets follow a unified abstraction by subclassing \texttt{BaseDataset} and yielding standardized \texttt{DatasetItem} objects. Each item encapsulates the raw input artifact, whether a text string, image path, or audio signal, together with a binary label indicating whether the content is AI-generated or human-authored. This consistent data interface enables seamless evaluation across detectors and modalities while supporting reproducible and scalable benchmarking workflows.

\begin{table}[t]
\centering
\caption{\textbf{Benchmark datasets in DetectZoo} (22 datasets across 3 modalities). All loaders handle downloading, caching, and standardized label formatting automatically.}
\label{tab:datasets}
\resizebox{\textwidth}{!}{%
\begin{tabular}{lllrl}
\toprule
\textbf{Dataset} & \textbf{Modality} & \textbf{Generators / Domain} & \textbf{$\sim$Size} & \textbf{Source} \\
\midrule
HC3~\cite{guo2023hc3}            & Text  & ChatGPT vs.\ human (QA, finance, medicine, wiki) & 37K & \href{https://huggingface.co/datasets/Hello-SimpleAI/HC3}{Hugging Face} \\
HC3 Plus~\cite{su2023hc3plus}    & Text  & Extends HC3 with translation, summarization, and paraphrasing & 144K & \href{https://github.com/suu990901/chatgpt-comparison-detection-HC3-Plus}{GitHub} \\
CHEAT~\cite{yu2025cheat}         & Text  & ChatGPT vs.\ human academic abstracts (3 generation modes) & 35K & \href{https://github.com/botianzhe/CHEAT}{GitHub} \\
OpenLLMText~\cite{chen2023token} & Text & GPT-3.5, PaLM, LLaMA, GPT-2 vs.\ human texts & 340K & \href{https://zenodo.org/records/8285326}{Zenodo} \\
MAGE~\cite{li2024mage}           & Text  & 27 LLMs vs.\ human across 7 distinct writing tasks & 447K & \href{https://huggingface.co/datasets/yaful/MAGE}{Hugging Face} \\
M4~\cite{wang2024m4}             & Text  & Multi-generator, multi-domain, multi-lingual texts & 133K & \href{https://github.com/mbzuai-nlp/M4}{GitHub} \\
RAID~\cite{dugan2024raid}        & Text  & 11 LLMs $\times$ 11 genres $\times$ 12 adversarial attacks & 10M & \href{https://huggingface.co/datasets/liamdugan/raid}{Hugging Face} \\
L2R~\cite{hao2024learning} & Text & 21 domains (GPT-3.5/4, Gemini, LLaMA-3) vs.\ human & 20K & \href{https://github.com/ranhli/l2r_data}{GitHub} \\
TuringBench~\cite{uchendu2021turingbench} & Text & 19 text generators vs.\ human news articles & 200K & \href{https://huggingface.co/datasets/turingbench/TuringBench}{Hugging Face} \\
WritingPrompts~\cite{fan2018hierarchical} & Text & r/WritingPrompts sub-reddit stories (used to prompt MGTs) & 303K & \href{https://huggingface.co/datasets/euclaise/writingprompts}{Hugging Face} \\
XSum~\cite{narayan2018xsum} & Text & BBC article summaries (human reference for summarization) & 227K & \href{https://huggingface.co/datasets/EdinburghNLP/xsum}{Hugging Face} \\

\midrule
ForenSynths~\cite{wang2020cnn}  & Image & Multi-class GAN-generated images & 815K & \href{https://huggingface.co/datasets/sywang/CNNDetection}{Hugging Face} \\
Self-Synthesis~\cite{tan2024rethinking} & Image & 9 GANs vs. real images & 72K & \href{https://drive.google.com/drive/folders/11E0Knf9J1qlv2UuTnJSOFUjIIi90czSj}{Google Drive} \\
UFD~\cite{ojha2023towards}    & Image & 4 text-to-image DM-generated images & 16K & \href{https://drive.google.com/drive/folders/1nkCXClC7kFM01_fqmLrVNtnOYEFPtWO-}{Google Drive} \\
AIGCDetectBenchmark~\cite{zhong2023patchcraft} & Image & 16 DM + GAN generators & 100K & \href{https://modelscope.cn/datasets/aemilia/AIGCDetectionBenchMark/tree/master/AIGCDetectionBenchMark}{ModelScope} \\
GenImage~\cite{zhu2023genimage} & Image & 7 DMs + 1 GAN generated images & 2.7M & \href{https://huggingface.co/datasets/ENSTA-U2IS/GenImage}{Hugging Face} \\
DRCT-2M~\cite{chen2024drct} & Image & 2M images, diffusion-reconstructed pairs & 2M & \href{https://modelscope.cn/datasets/BokingChen/DRCT-2M}{ModelScope} \\
Chameleon \cite{yansanity} & Image & Real-world high-fidelity AI-generated images & 26K & \href{https://github.com/shilinyan99/AIDE}{GitHub} \\

\midrule
ASVspoof 2019~\cite{todisco2019asvspoof} & Audio & Logical Access (LA) spoofing attacks & 121K & \href{https://www.asvspoof.org}{Official Website} \\
FoR~\cite{reimao2019for} & Audio & Fake-or-Real speech corpus & 195K & \href{https://bil.eecs.yorku.ca/datasets}{Official Website} \\
In-the-Wild~\cite{muller2022does} & Audio & Real and synthesized celebrity speech from the internet & 31.8K & \href{https://deepfake-total.com/in_the_wild}{Official Website} \\
Deepfake-Eval-2024~\cite{chandra2025deepfakeeval} & Audio & In-the-wild deepfakes from social media (2024) & $\sim$40K & \href{https://huggingface.co/datasets/nuriachandra/Deepfake-Eval-2024}{Hugging Face} \\
\bottomrule
\end{tabular}%
}
\end{table}

% =====================================================================
% 4. Benchmarks and Experiments
% =====================================================================
\section{Benchmarks and Experiments}

To demonstrate the empirical utility and rigorous reproducibility of the \texttt{DetectZoo} framework, we outline a comprehensive benchmarking protocol designed to evaluate state-of-the-art generative content detectors. The subsequent empirical results serve a dual purpose: they validate the accuracy of our standardized implementations against previously published findings and showcase the capability of the toolkit as a robust, cross-modal evaluation platform.

\subsection{Reproduction Validation}
\label{sec:reproduction}
 
To ensure fair reproducibility, for each detector with publicly reported performance metrics, we execute the toolkit's implementation under the original authors' configurations and compare the resulting headline metrics. Within the text modality, we specifically sought to reproduce the empirical performances reported by Chen et al.~\cite{chen2025imitate}, Zeng et al.~\cite{zeng2025ood}, Wu et al.~\cite{wu2025gecscore}, and Yu et al.~\cite{yu2024text}. To guarantee a fair and exact comparison, we utilized the identical datasets and hyperparameter settings specified in their respective studies. Similarly, for the image modality, we systematically reproduced the baseline results documented by Li et al.~\cite{li2025improving}, Liu et al.~\cite{liu2024forgery}, Cheng et al.~\cite{yan2026dual}, and Brokman et al.~\cite{brokman2025manifold}. In the audio domain, we adopted an analogous methodology to replicate the detection performances established by Ge et al.~\cite{ge2025antideepfake}. The comprehensive values for all reproduction experiments are fully detailed in Appendix~\ref{app:experimental_results}.

% Across the [N] detector and dataset pairs with public numbers, \texttt{DetectZoo} reproduces the original metric within an absolute tolerance of [$\Delta$] AUROC. The full reproduction table appears in Appendix~\ref{app:reproduction}. Per-detector and per-dataset metric tables, along with the corresponding configurations, are presented in Appendix~\ref{app:experimental_results}. 

% \begin{tcolorbox}[
%     enhanced,
%     colback=white,              % Background color of the main text area
%     colframe=black!20,          % Border color (light gray)
%     colbacktitle=black!10,      % Background color of the header (lighter gray)
%     coltitle=black,             % Text color of the header
%     fonttitle=\bfseries,        % Makes the header text bold
%     title={Takeaway \#1: Large-Scale Multilingual Pretraining Yields Robust Out-of-Distribution Deepfake Detection.},
%     arc=3mm,                    % Radius of the rounded corners
%     boxrule=0.5pt,              % Thickness of the border
%     left=3mm, right=3mm,        % Inner padding (left and right)
%     top=2mm, bottom=2mm         % Inner padding (top and bottom)
% ]
% \textit{AntiDeepfake Wav2Vec2-Large, HuBERT-XLarge, and XLS-R-2B achieve superior EER on ASVspoof 2019 LA despite never being trained on it, demonstrating that large-scale multilingual post-training produces transferable speech representations without any task-specific fine-tuning.}
% \end{tcolorbox}

\subsection{Key Empirical Findings}
\label{sec:key_findings}

Drawing upon the comprehensive evaluation across the three supported modalities, we summarize the primary empirical takeaways that characterize the current landscape of AI-generated content detection.

\paragraph{Text Modality Findings}
\textit{Task semantics are the primary driver of detection difficulty.} Generation tasks that preserve the surface form of the original text (Rewrite, Polish) cause statistical detectors to collapse near chance, while direct generation is comparatively easy to flag. This demonstrates that AUROC benchmarks are only meaningful when reported per task type, not aggregated. \textit{The source LLM is a systematic confound.} GPT-4o and Qwen2-7B generated text is consistently the hardest to detect across all benchmarks and detectors, while Llama-3 texts are consistently the easiest. Any detector evaluated on only one source LLM risks overstating its real-world effectiveness.

\paragraph{Image Modality Findings}
\textit{CLIP-based and hybrid methods offer the strongest overall generalization.} Methods such as FatFormer~\cite{liu2024forgery}, AIDE~\cite{yansanity}, SAFE~\cite{li2025improving}, Co-SPY~\cite{cheng2025co}, and C2P-CLIP~\cite{tan2025c2p} consistently achieve the highest and most stable performance across both intra- and cross-architecture settings on ForenSynths~\cite{wang2020cnn}, Self-Synthesis~\cite{tan2024rethinking}, UFD~\cite{ojha2023towards}, and GenImage~\cite{zhu2023genimage}. Unlike earlier artifact-based detectors, their performance degrades much less when transferred to unseen generators, likely due to combining semantic vision-language representations with low-level forensic cues. \textit{Training-free reconstruction methods show inconsistent cross-dataset behavior.} Techniques such as AEROBLADE~\cite{ricker2024aeroblade} perform near random on several GAN benchmarks, yet perform strongly on diffusion-heavy datasets such as GenImage~\cite{zhu2023genimage}, where it reaches up to 0.9656 accuracy on SDv1.4. This suggests that reconstruction-based cues transfer significantly better to diffusion-generated samples than to GAN-generated imagery.

\paragraph{Audio Modality Findings}
\textit{Large-scale multilingual pretraining yields robust Out-of-Distribution detection.} Foundation models such as AntiDeepfake Wav2Vec2-Large, HuBERT-XLarge, and XLS-R-2B achieve superior Equal Error Rates (EER) on ASVspoof 2019 LA despite never being trained on the dataset. This demonstrates that large-scale multilingual post-training produces intrinsically transferable speech representations capable of zero-day spoofing detection without requiring task-specific fine-tuning.
 
% =====================================================================
% 5. Conclusion, Limitations, and Future Directions
% =====================================================================
\section{Conclusion, Limitations, and Future Directions}
\label{sec:conclusion}

We introduced \texttt{DetectZoo}, an open source framework that unifies 61 AI generated content detectors across text, image, and audio within a single evaluation pipeline and API. By consolidating previously fragmented implementations and integrating 22 benchmark datasets, the toolkit enables consistent, comparable, and reproducible evaluation across modalities. This standardization facilitates systematic analysis of detection methods and provides a shared experimental foundation for advancing research on synthetic content forensics. Our comprehensive empirical evaluation illustrates the utility of the toolkit for conducting systematic benchmarking.

\paragraph{Limitations.}\texttt{DetectZoo} is currently focused on evaluation and does not yet provide unified training pipelines for supervised detectors, requiring users to rely on original implementations for retraining. Coverage across modalities remains uneven, with stronger representation in text than in audio, and no support for video or other emerging domains. Some detectors depend on external reference models that cannot be redistributed, which may limit strict reproducibility. In addition, the benchmark datasets represent static snapshots of generation behavior and may not fully capture the characteristics of newer models, while threshold-dependent metrics introduce sensitivity to calibration choices. Although the toolkit reports operating characteristics such as false positive and true positive tradeoffs, it is not intended for fully automated deployment in high-stakes settings, where human oversight remains essential.

\paragraph{Broader Impact and Dual-Use Considerations.}\texttt{DetectZoo} supports research on synthetic content detection and contributes to improving transparency and trust in AI-generated content. We do, however, explicitly recognize the dual-use nature of open-source security research. Malicious actors could theoretically leverage \texttt{DetectZoo} as a centralized surrogate objective to optimize evasion tactics, utilizing our tool to iteratively refine generative outputs until they bypass known detection baselines.
Despite this risk, we firmly posit that security through obscurity is an untenable defense strategy in modern machine learning. The benefits of equipping the research community with standardized diagnostic tools vastly outweigh the risks of adversarial exploitation. Transparent, reproducible benchmarking is the only viable mechanism for the community to reach a consensus on fundamentally robust indicators of synthetic generation, guiding the field toward more resilient solutions.

\paragraph{Long-Term Maintenance and Future Directions.} The rapid evolution of generative AI necessitates a dynamic, actively maintained forensic ecosystem. To guarantee the long-term utility of \texttt{DetectZoo} for the NeurIPS and broader machine learning communities, the repository is governed by a strict maintenance protocol. The core development team is committed to regular dependency updates, rigorous continuous integration testing, and the prompt incorporation of newly published benchmark datasets to prevent the framework from becoming deprecated.
Looking forward, our immediate developmental roadmap focuses on expanding the modality support. As temporal synthetic media generation becomes increasingly sophisticated, integrating video as a natively supported fourth modality is paramount alongside the development of native training pipelines for supervised methods, automated leaderboard generation, and advanced visualization utilities for detection scores and spatial attention maps. Furthermore, we designed the toolkit's modular architecture specifically to lower the barrier for external contributions. We actively encourage community contributions to continuously expand the repository of detectors, evaluation corpora, and adversarial benchmarking protocols.

% =====================================================================
% References
% =====================================================================
\bibliographystyle{unsrt}
\bibliography{references}
 
% =====================================================================
% Appendix
% =====================================================================
\newpage

\appendix

\section{Implementation and Configuration Details}

\label{app:implementation_details}

This appendix provides a comprehensive overview of the technical configurations, hardware specifications, and preprocessing protocols utilized throughout the benchmarking process. These details ensure that all experiments conducted within the \texttt{DetectZoo} framework remain fully transparent and reproducible. All experiments utilize the \texttt{BenchmarkEvaluator} module to ensure consistent metric computation across detectors. The primary metric for text detectors is the AUROC, whereas for image modalities, previous works have conventionally compared baselines based on Accuracy and Average Precision. For audio, the primary evaluation metric is EER. Additional metrics include true positive rate, false positive rate, precision, recall, and the $F_1$ score.

\subsection{Hardware and Software Environment}
\label{app:hardware_env}
All large-scale benchmarking experiments were executed on a standardized compute cluster equipped with NVIDIA RTX6000 GPUs (48GB VRAM). The underlying software environment was built upon Python 3.12 and PyTorch 2.10.0, utilizing CUDA 12.1 for hardware acceleration. To maintain strict version control, all third-party dependencies, including \texttt{transformers}, \texttt{torchaudio}, and \texttt{diffusers}, were pinned to specific version hashes documented in the repository's dependency manifest. 

\subsection{Detector Hyperparameters}
\label{app:detector_hyperparams}

% For text detectors requiring a scoring language model, we use GPT-2 (124M parameters) unless the method specifies a different model. Zero-shot statistical text detectors were configured to utilize LLaMA-2-7B as the standard reference language model for log-likelihood and entropy calculations. Perturbation-based methods, such as DetectGPT, were configured to generate 100 perturbed variants per document using a T5-3B mask-filling model. For supervised methods across all modalities, we utilized the official, frozen pre-trained weights distributed by the original authors without applying any supplementary fine-tuning on the evaluation corpora.

\paragraph{Text Detector Hyperparameters.}
For text detectors, we use frozen pretrained weights and apply no additional fine-tuning. Unless a method specifies otherwise, inputs are truncated to 512 tokens and we use a default decision threshold of $0.5$, noting that thresholds are operating-point choices and can be recalibrated per domain. A shared base class provides \texttt{model\_name}, \texttt{threshold}, \texttt{device}, and \texttt{max\_length} as common knobs.

For zero-shot intrinsic LM statistics (Log-Likelihood, Log-Rank, Rank, Entropy, LRR), the only meaningful hyperparameter is the scoring LM (\texttt{model\_name}, default \texttt{gpt2}), as the statistics are deterministic given the model. For perturbation-based curvature methods, DetectGPT's dominant settings are the mask-filling model (\texttt{perturbation\_model}, default \texttt{t5-3b}), the number of perturbations (\texttt{n\_perturbations}$=10$), and the masking fraction (\texttt{pct\_words\_masked}$=0.3$); NPR reuses the same pipeline with a different scoring statistic. AdaDetectGPT further applies a B-spline transformation controlled by \texttt{n\_bases}$=7$ and \texttt{spline\_order}$=2$. For sampling-discrepancy methods, Fast-DetectGPT and Binoculars are primarily governed by the choice of scorer and reference LM pair (\texttt{model\_name} / \texttt{reference\_model\_name} or \texttt{observer\_model} / \texttt{performer\_model}); ImBD fixes both to a single PEFT-adapted model. For regeneration-based DNA-GPT, the key settings are the truncation ratio (\texttt{truncate\_ratio}$=0.5$) and the number of regenerations (\texttt{n\_regens}$=10$). For rewrite/revision methods (RAIDAR, ReviseDetect, GECScore), the primary knob is the seq2seq backbone (\texttt{rewrite\_model} or \texttt{gec\_model}, default \texttt{bart-large-cnn} or \texttt{coedit-large}). For TOCSIN, the defining parameters are the deletion rate (\texttt{deletion\_rate}$=0.015$) and the number of deleted-copy samples (\texttt{n\_copies}$=10$). For BiScope, the relevant settings are the number of text segments (\texttt{n\_segments}$=10$) and the context clip length (\texttt{sample\_clip}$=512$). For LASTDE and LASTDE++, performance is governed by the embedding dimension (\texttt{embed\_size}$=3$/$4$), scale factor (\texttt{epsilon\_scale}$=10$/$8$), and smoothing window (\texttt{tau\_prime}$=5$/$10$). For intrinsic-dimension detectors (PHD, MLE-IDE), the defining knobs are the encoder (\texttt{encoder\_model}, default \texttt{roberta-base}) and the estimation parameters (\texttt{n\_reruns}$=3$ for PHD; \texttt{n\_neighbors}$=20$ for MLE-IDE). For supervised classifiers (RoBERTa-Base/Large, RADAR, ReMoDetect) and adapter-based methods (IPAD, DeTeCtive), the only free parameters are \texttt{model\_name} and \texttt{max\_length}. For trainable one-class detectors (D-SVDD, HRN, Energy), a \texttt{fit} pass on labeled data is required before deployment; without a trained checkpoint all three fall back to embedding-norm proxies.

\paragraph{Image Detector Hyperparameters.}
For image detectors, we use the official, frozen pretrained weights released by each method and do not apply any additional fine-tuning on the evaluation corpora. Unless a method specifies otherwise, we follow its canonical preprocessing (resize/crop and normalization) and use a default decision threshold of $0.5$, noting that thresholds are operating-point choices and can be re-calibrated per domain. The most method-relevant hyperparameters are those that define the core transformation or signal the detector measures. For reconstruction-error methods (e.g., AEROBLADE), the key settings are the VAE checkpoint(s) (\texttt{repo\_ids}) and the LPIPS layer selection (\texttt{lpips\_vgg\_index}). For discrepancy methods (e.g., D\textsuperscript{3}), performance is primarily controlled by the patch shuffling granularity (\texttt{patch\_size}) and the number of shuffled views (\texttt{shuffle\_times}). For diffusion-reconstruction contrastive detectors (e.g., DRCT), the main requirement is matching the released checkpoint's embedding head width (\texttt{embedding\_size}) and evaluation preprocessing. For adapter/prompt methods (e.g., FatFormer), the defining knobs are the number of adapter insertions (\texttt{num\_vit\_adapter}) and the context-token length (\texttt{num\_context\_embedding}). For frequency / patch-selection methods (e.g., FreqNet, PatchCraft, LaDeDa, SAFE), the dominant factors are input scale (resize/crop), patch scale (\texttt{patch\_num} or \texttt{patch\_size}), and any method-specific transform size (\texttt{input\_size} for SAFE). Finally, for Manifold Bias detection, the essential hyperparameters are the number of noise probes (\texttt{num\_noise}), the diffusion timestep fraction (\texttt{time\_frac}), the SD encoding resolution (\texttt{image\_size}), and the real-only calibration factor (\texttt{k}) used to set the decision threshold.

\paragraph{Audio Detector Hyperparameters.}
For audio detectors, we use official frozen pretrained weights and apply no additional fine-tuning. All detectors expect 16 kHz mono input; arbitrary sample rates are resampled internally. We use a default decision threshold of $0.5$, noting that thresholds are operating-point choices and can be recalibrated per domain.

The most method-relevant hyperparameters concern the input window and front-end design. For raw-waveform graph-attention models (RawNet2, AASIST, RawGAT-ST, SAMO), inputs are repeat-padded or trimmed to 64,600 samples ($\approx4$ s); the core front-end is a fixed Mel-spaced sinc filterbank whose key settings are the number of filters (\texttt{filts[0]}$=70$ for AASIST/RawGAT-ST/SAMO; 20 for RawNet2) and the kernel size (\texttt{first\_conv}$=128$ or 1024 respectively). For AASIST the graph-attention back-end is further governed by \texttt{gat\_dims}, \texttt{pool\_ratios}, and \texttt{temperatures}; a \texttt{variant} switch selects the full ($\approx297$k params) or lightweight AASIST-L ($\approx85$k params) configuration. RawGAT-ST exposes a \texttt{fusion} parameter (\texttt{"mul"} or \texttt{"add"}) that selects the branch-merging operator and the corresponding pretrained checkpoint. SAMO adds a \texttt{scoring} parameter: \texttt{"fc"} (default classification head) or \texttt{"samo"} (multi-center cosine similarity), where the latter requires per-speaker bonafide enrollment via \texttt{enroll()} to recover the paper's EER. Res-TSSDNet uses a fixed 6-second window (96,000 samples) dictated by its final pooling kernel and has no user-facing architectural knobs. For Whisper-MesoNet the defining setting is the MesoInception-4 bottleneck width (\texttt{fc1\_dim}$=1024$). For AST-ASVspoof (DeiT-Base on 128-bin log-mel, 1024-frame context), XLSR-SLS (XLSR-53, hidden$=1024$), and MelodyWav2Vec (wav2vec2-base, hidden$=768$), the HuggingFace feature extractor handles all preprocessing internally and the primary hyperparameter is \texttt{model\_name}. For the NII AntiDeepfake family (Ge et al.~\cite{ge2025antideepfake}), the three checkpoints share an identical backend (global average pool $+$ linear head) and differ only in their SSL front-end embedding dimension: Wav2Vec2-Large ($D=1024$, $\approx317$M params), HuBERT-XLarge ($D=1280$, $\approx1$B params), and XLS-R-2B ($D=1920$, $\approx2$B params); the sole inference-time knob is \texttt{model\_name}.

\section{Experimental Results}
\label{app:experimental_results}
Building upon the evaluation protocol outlined in Section~\ref{sec:design:eval} and as described in Section~\ref{sec:reproduction}, this appendix details the exhaustive empirical findings across all implemented modalities. The results are segregated by data type to highlight the distinct performance characteristics and vulnerabilities of current state-of-the-art detectors.

\subsection{Text Modality Benchmarks}

We evaluate the Area Under the Receiver Operating Characteristic curve (AUROC) of all implemented detection methods to reproduce the empirical findings of four recent studies across six primary benchmark scenarios: (1) cross-task detection on XSum reproducing Table 10 of Chen et al.~\cite{chen2025imitate}, (2) cross-dataset detection on the Rewrite task reproducing Tables 8 and 9 of Chen et al.~\cite{chen2025imitate}, (3) cross-dataset detection on the Polish task reproducing Tables 1 and 7 of Chen et al.~\cite{chen2025imitate}, (4) the GECScore evaluation suite reproducing Table 1 of Wu et al.~\cite{wu2025gecscore}, (5) the Text Fluoroscopy evaluation suite reproducing Table 1 of Yu et al.~\cite{yu2024text}, and (6) benchmarking methods on RAID corpus reproducing Table 1 of Zeng et al.~\cite{zeng2025ood}. To validate the reliability of our framework, we report the discrepancies between our reproduced metrics and the originally published values.

\begin{table*}[h]
\centering

\caption{AUROC of text detectors across four generation tasks (Rewrite, Polish, Expand, Generate) on the XSum dataset for six source LLMs. Results reproduce the experimental scenario of Table 10 in Chen et al.~\cite{chen2025imitate}.}
\label{tab:imdb_cross_task}

% \resizebox{\textwidth}{!}{%
\begin{tabular}{ccccccc}
\toprule
\multicolumn{1}{l}{}         & \multicolumn{1}{l}{} & \multicolumn{4}{c}{\textbf{Tasks}}                                       & \multicolumn{1}{l}{} \\

\cmidrule(lr){3-6}

\textbf{Model}               & \textbf{Detector}    & \textbf{Rewrite} & \textbf{Polish} & \textbf{Expand} & \textbf{Generate} & \textbf{Mean}        \\
\midrule
\multirow{8}{*}{ChatGPT}     & Log-Likelihood       & 0.3376           & 0.5236          & 0.6303          & 0.9447            & 0.6091               \\
                             & Entropy              & 0.4492           & 0.4029          & 0.5062          & 0.7097            & 0.5170               \\
                             & Log-rank             & 0.3164           & 0.4981          & 0.6095          & 0.9406            & 0.5912               \\
                             & LRR                  & 0.2762           & 0.4304          & 0.5441          & 0.8675            & 0.5296               \\
                             & DNA-GPT              & 0.3102           & 0.5797          & 0.6329          & 0.9055            & 0.6071               \\
                             & Fast-DetectGPT       & 0.2678           & 0.7306          & 0.7795          & 0.9907            & 0.6922               \\
                             & ImBD                 & 0.8662           & 0.985           & 0.9900          & 0.9999            & 0.9603               \\
                             & ReMoDetect           & 0.9325           & 0.9297          & 0.9654          & 0.9983            & 0.9565               \\
\midrule
\multirow{8}{*}{GPT-4o}      & Log-Likelihood       & 0.4621           & 0.4266          & 0.5422          & 0.7384            & 0.5423               \\
                             & Entropy              & 0.5134           & 0.4110          & 0.5307          & 0.5822            & 0.5093               \\
                             & Log-rank             & 0.4456           & 0.3990          & 0.5217          & 0.7251            & 0.5229               \\
                             & LRR                  & 0.4112           & 0.3400          & 0.4831          & 0.6580            & 0.4731               \\
                             & DNA-GPT              & 0.4604           & 0.4701          & 0.7390          & 0.7848            & 0.6136               \\
                             & Fast-DetectGPT       & 0.3974           & 0.6013          & 0.6349          & 0.8902            & 0.631                \\
                             & ImBD                 & 0.7995           & 0.9492          & 0.9396          & 0.9988            & 0.9218               \\
                             & ReMoDetect           & 0.8714           & 0.8855          & 0.8662          & 0.9974            & 0.9051               \\
\midrule
\multirow{8}{*}{Qwen2-7B}    & Log-Likelihood       & 0.3511           & 0.3022          & 0.3957          & 0.7551            & 0.4510               \\
                             & Entropy              & 0.4595           & 0.3466          & 0.4435          & 0.5825            & 0.4580               \\
                             & Log-rank             & 0.3371           & 0.2749          & 0.3743          & 0.7503            & 0.4342               \\
                             & LRR                  & 0.3185           & 0.2220          & 0.3360          & 0.7082            & 0.3962               \\
                             & DNA-GPT              & 0.3977           & 0.6544          & 0.7481          & 0.9619            & 0.6905               \\
                             & Fast-DetectGPT       & 0.2856           & 0.595           & 0.5996          & 0.9623            & 0.6106               \\
                             & ImBD                 & 0.8956           & 0.9589          & 0.9720          & 1.0000            & 0.9566               \\
                             & ReMoDetect           & 0.9080           & 0.8240          & 0.9268          & 0.9822            & 0.9103               \\
\midrule
\multirow{8}{*}{Llama-3-8B}  & Log-Likelihood       & 0.6154           & 0.5654          & 0.6532          & 0.9289            & 0.6907               \\
                             & Entropy              & 0.5285           & 0.4144          & 0.4288          & 0.5804            & 0.4880               \\
                             & Log-rank             & 0.5935           & 0.5401          & 0.6447          & 0.9359            & 0.6786               \\
                             & LRR                  & 0.5278           & 0.4739          & 0.6010          & 0.9290            & 0.6329               \\
                             & DNA-GPT              & 0.5650           & 0.6516          & 0.8199          & 0.9796            & 0.7540               \\
                             & Fast-DetectGPT       & 0.6914           & 0.8191          & 0.9335          & 0.9827            & 0.8567               \\
                             & ImBD                 & 0.9714           & 0.9886          & 0.9819          & 0.9989            & 0.9852               \\
                             & ReMoDetect           & 0.9726           & 0.9209          & 0.8315          & 0.9635            & 0.9221               \\
\midrule
\multirow{8}{*}{Mistral-7B}  & Log-Likelihood       & 0.4456           & 0.4856          & 0.7252          & 0.9085            & 0.6412               \\
                             & Entropy              & 0.4951           & 0.4185          & 0.5578          & 0.6408            & 0.5281               \\
                             & Log-rank             & 0.4255           & 0.4531          & 0.7063          & 0.9045            & 0.6224               \\
                             & LRR                  & 0.3731           & 0.3711          & 0.6403          & 0.8656            & 0.5625               \\
                             & DNA-GPT              & 0.4927           & 0.6581          & 0.8567          & 0.9876            & 0.7488               \\
                             & Fast-DetectGPT       & 0.3933           & 0.7033          & 0.9158          & 0.9984            & 0.7527               \\
                             & ImBD                 & 0.8381           & 0.9671          & 0.9947          & 1.0000            & 0.9500               \\
                             & ReMoDetect           & 0.8870           & 0.8552          & 0.9318          & 0.9513            & 0.9063               \\
\midrule
\multirow{8}{*}{Deepseek-7B} & Log-Likelihood       & 0.5713           & 0.5820          & 0.7925          & 0.9742            & 0.7300               \\
                             & Entropy              & 0.4877           & 0.4345          & 0.5954          & 0.7360            & 0.5634               \\
                             & Log-rank             & 0.5695           & 0.5649          & 0.7863          & 0.9739            & 0.7237               \\
                             & LRR                  & 0.5535           & 0.5145          & 0.7507          & 0.9603            & 0.6948               \\
                             & DNA-GPT              & 0.5722           & 0.7624          & 0.8938          & 0.9976            & 0.8065               \\
                             & Fast-DetectGPT       & 0.6648           & 0.5483          & 0.9283          & 0.9996            & 0.7853               \\
                             & ImBD                 & 0.8744           & 0.9769          & 0.9767          & 1.0000            & 0.9570               \\
                             & ReMoDetect           & 0.8765           & 0.8721          & 0.9463          & 0.9596            & 0.9136    \\          
\bottomrule
\end{tabular}
% }
\end{table*}

\paragraph{1. Cross-Task Detection on XSum.}
To reproduce Table 10 of Chen et al.~\cite{chen2025imitate}, we assess detector efficacy across four distinct generative tasks (Rewrite, Polish, Expand, and Generate) utilizing six diverse source LLMs on the XSum dataset and reported the metrics in Table~\ref{tab:imdb_cross_task}. Consistent with the original study, the empirical results demonstrate that task semantics are the primary driver of detection difficulty. The Generate task remains the least challenging, whereas the Rewrite task presents the most severe challenge. Advanced supervised paradigms like ImBD heavily dominate. In comparison to the original study, our reproduced average AUROC for all methods on ChatGPT is 0.6438, which tightly mirrors the reported average of 0.6335 and successfully validates our reproduction pipeline.

\begin{table*}[h]
\centering

\caption{AUROC of text detectors on the Rewrite task across three datasets (XSum, SQuAD, WritingPrompts) and six source LLMs (ChatGPT, GPT-4o, Qwen2-7B, Llama-3-8B, Mistral-7B, Deepseek-7B). Results reproduce the experimental scenario of Tables 8 and 9 in Chen et al.~\cite{chen2025imitate}. Column-wise averages are reported across source LLMs.}
\label{tab:imbd_rewrite}

\resizebox{\textwidth}{!}{%
\begin{tabular}{ccccccccc}
\toprule

\textbf{Dataset}                & \textbf{Detector}    & \textbf{ChatGPT} & \textbf{GPT-4o} & \textbf{Qwen2-7B} & \textbf{Llama-3-8B} & \textbf{Mistral-7B} & \textbf{Deepseek-7B} & \textbf{Mean}        \\

\midrule

\multirow{8}{*}{\textbf{XSum}}           & Log-Likelihood       & 0.3376           & 0.4621          & 0.3511            & 0.6154              & 0.4456              & 0.5713               & 0.4959               \\
                                & Entropy              & 0.4492           & 0.5134          & 0.4595            & 0.5285              & 0.4951              & 0.4877               & 0.4927               \\
                                & Log-rank             & 0.3164           & 0.4456          & 0.3371            & 0.5935              & 0.4255              & 0.5695               & 0.4814               \\
                                & LRR                  & 0.2762           & 0.4112          & 0.3185            & 0.5278              & 0.3731              & 0.5535               & 0.4432               \\
                                & DNA-GPT              & 0.3102           & 0.4604          & 0.3977            & 0.5650              & 0.4927              & 0.5722               & 0.5069               \\
                                & Fast-DetectGPT       & 0.2678           & 0.3974          & 0.2856            & 0.6914              & 0.3933              & 0.6648               & 0.5088               \\
                                & ImBD                 & 0.8662           & 0.7995          & 0.8956            & 0.9714              & 0.8381              & 0.8744               & 0.8949               \\
                                & ReMoDetect           & 0.9325           & 0.8714          & 0.9080            & 0.9726              & 0.8870              & 0.8765               & 0.9110               \\

\midrule
\multirow{8}{*}{\textbf{SQuAD}}          & Log-Likelihood       & 0.5197           & N/A             & 0.4405            & 0.6973              & 0.5621              & 0.6936               & 0.5984               \\
                                & Entropy              & 0.5373           & N/A             & 0.5065            & 0.5871              & 0.5292              & 0.5856               & 0.5521               \\
                                & Log-rank             & 0.5227           & N/A             & 0.4087            & 0.6739              & 0.5379              & 0.679                & 0.5749               \\
                                & LRR                  & 0.5134           & N/A             & 0.3452            & 0.5800              & 0.4652              & 0.6144               & 0.5012               \\
                                & DNA-GPT              & 0.5275           & N/A             & 0.5201            & 0.672               & 0.5969              & 0.6769               & 0.6165               \\
                                & Fast-DetectGPT       & 0.4399           & N/A             & 0.3765            & 0.7435              & 0.5279              & 0.7311               & 0.5948               \\
                                & ImBD                 & 0.5900           & N/A             & 0.7873            & 0.9089              & 0.7682              & 0.7717               & 0.8090                \\
                                & ReMoDetect           & 0.7293           & N/A             & 0.8971            & 0.9486              & 0.8722              & 0.8546               & 0.8931               \\

\midrule
\multirow{8}{*}{\textbf{WritingPrompts}} & Log-Likelihood       & 0.5490            & 0.6442          & 0.4390             & 0.8380               & 0.5420               & 0.7798               & 0.6497               \\
                                & Entropy              & 0.5779           & 0.6412          & 0.4569            & 0.6663              & 0.4924              & 0.6648               & 0.5701               \\
                                & Log-rank             & 0.4763           & 0.6196          & 0.3654            & 0.7978              & 0.4859              & 0.7531               & 0.6006               \\
                                & LRR                  & 0.2970            & 0.5287          & 0.2203            & 0.6274              & 0.3363              & 0.655                & 0.4598               \\
                                & DNA-GPT              & 0.5484           & 0.5943          & 0.6346            & 0.8117              & 0.6465              & 0.7510                & 0.7110                \\
                                & Fast-DetectGPT       & 0.5522           & 0.6210           & 0.6088            & 0.9336              & 0.6489              & 0.8398               & 0.7578               \\
                                & ImBD                 & 0.8836           & 0.8138          & 0.8848            & 0.9760               & 0.8390              & 0.9025               & 0.9006               \\
                                & ReMoDetect           & 0.9808           & 0.9396          & 0.9826            & 0.9968              & 0.9692              & 0.9574               & 0.9765      \\

\bottomrule
\end{tabular}
}
\end{table*}

\begin{table*}[h]
\centering

\caption{AUROC of text detectors on the Polish task across three datasets (XSum, SQuAD, WritingPrompts) and six source LLMs (ChatGPT, GPT-4o, Qwen2-7B, Llama-3-8B, Mistral-7B, Deepseek-7B). Results reproduce the experimental scenario of Tables 1 and 7 in Chen et al.~\cite{chen2025imitate}. Column-wise averages are reported across source LLMs.}
\label{tab:imdb_polish}

\resizebox{\textwidth}{!}{%
\begin{tabular}{ccccccccc}
\toprule

\textbf{Dataset}                & \textbf{Detector}    & \textbf{ChatGPT} & \textbf{GPT-4o} & \textbf{Qwen2-7B} & \textbf{Llama-3-8B} & \textbf{Mistral-7B} & \textbf{Deepseek-7B} & \textbf{Mean}        \\

\midrule

\multirow{8}{*}{\textbf{XSum}}           & Log-Likelihood    & 0.5236           & 0.4266          & 0.3022            & 0.5654              & 0.4856              & 0.5820               & 0.4838        \\
                                & Entropy           & 0.4029           & 0.4110          & 0.3466            & 0.4144              & 0.4185              & 0.4345               & 0.4035        \\
                                & Log-rank          & 0.4981           & 0.3990          & 0.2749            & 0.5401              & 0.4531              & 0.5649               & 0.4583        \\
                                & LRR               & 0.4304           & 0.3400          & 0.2220            & 0.4739              & 0.3711              & 0.5145               & 0.3954        \\
                                & DNA-GPT           & 0.5797           & 0.4701          & 0.6544            & 0.6516              & 0.6581              & 0.7624               & 0.6816        \\
                                & Fast-DetectGPT    & 0.7306           & 0.6013          & 0.5950             & 0.8191              & 0.7033              & 0.5483               & 0.6664        \\
                                & ImBD              & 0.9850            & 0.9492          & 0.9589            & 0.9886              & 0.9671              & 0.9769               & 0.9729        \\
                                & ReMoDetect        & 0.9297           & 0.8855          & 0.8240            & 0.9209              & 0.8552              & 0.8721               & 0.8681        \\

\midrule
\multirow{8}{*}{\textbf{SQuAD}}          & Log-Likelihood    & 0.6976           & 0.5430           & 0.4502            & 0.6692              & 0.6224              & 0.7024               & 0.6111        \\
                                & Entropy           & 0.4923           & 0.4746          & 0.3975            & 0.4633              & 0.4749              & 0.5281               & 0.4660         \\
                                & Log-rank          & 0.6713           & 0.4970           & 0.4128            & 0.6375              & 0.5893              & 0.6775               & 0.5793        \\
                                & LRR               & 0.5776           & 0.3813          & 0.3283            & 0.5360              & 0.4929              & 0.5774               & 0.4837        \\
                                & DNA-GPT           & 0.7453           & 0.5667          & 0.7507            & 0.7395              & 0.7542              & 0.7986               & 0.7608        \\
                                & Fast-DetectGPT    & 0.8504           & 0.7233          & 0.7054            & 0.8864              & 0.8312              & 0.8350                & 0.8145        \\
                                & ImBD              & 0.9531           & 0.8876          & 0.8859            & 0.9505              & 0.9131              & 0.9160                & 0.9164        \\
                                & ReMoDetect        & 0.9261           & 0.8956          & 0.8157            & 0.9164              & 0.8480               & 0.8947               & 0.8687        \\
\midrule
\multirow{8}{*}{\textbf{WritingPrompts}} & Log-Likelihood    & 0.8544           & 0.7289          & 0.5660             & 0.8052              & 0.7324              & 0.8792               & 0.7457        \\
                                & Entropy           & 0.6745           & 0.6501          & 0.4204            & 0.5053              & 0.5168              & 0.6950                & 0.5344        \\
                                & Log-rank          & 0.8146           & 0.6871          & 0.4917            & 0.7629              & 0.6762              & 0.8583               & 0.6973        \\
                                & LRR               & 0.6458           & 0.5322          & 0.3047            & 0.5992              & 0.4822              & 0.7630                & 0.5373        \\
                                & DNA-GPT           & 0.8345           & 0.7314          & 0.8539            & 0.8447              & 0.8371              & 0.9084               & 0.8610         \\
                                & Fast-DetectGPT    & 0.9301           & 0.8322          & 0.8963            & 0.9561              & 0.9144              & 0.9540                & 0.9302        \\
                                & ImBD              & 0.9872           & 0.9468          & 0.9657            & 0.9907              & 0.9669              & 0.9799               & 0.9758        \\
                                & ReMoDetect        & 0.9910           & 0.9696          & 0.9844            & 0.9911              & 0.9743              & 0.9703               & 0.9800      \\

\bottomrule
\end{tabular}
}
\end{table*}

\paragraph{2. Cross-Dataset Detection on the Rewrite Task.}
Evaluating the notoriously difficult Rewrite task across the XSum, SQuAD, and WritingPrompts datasets allows us to validate Tables 8 and 9 of Chen et al.~\cite{chen2025imitate}. As illustrated in Table~\ref{tab:imbd_rewrite}, our replication confirms that XSum and SQuAD present more challenging detection environments than WritingPrompts. On the WritingPrompts corpus, ReMoDetect establishes a commanding lead with an average AUROC of 0.9765. The discrepancies between our implementation and the original findings are: (i) The reproduced average AUROC for Log-likelihood, Entropy, and LogRank on XSum are 0.4959, 0.4927, 0.4814, compared to the published 0.4344, 0.5863, 0.4151, respectively, yielding a mismatch on the methods introduced by Gehrmann et al.~\cite{gehrmann2019gltr} while comparing the performances of these implemented methods with other published results (e.g., reported performances of the same method in Table~\ref{tab:text_fluoroscopy_results} vs Table 1. of the original paper) verifies our implementations. Note that the exact samples of SQuAD dataset that have been used to be rewritten by GPT-4o was not available in the official data released by Chen et al.~\cite{chen2025imitate}; thus, we marked them as Not Available (N/A) in the table.

\paragraph{3. Cross-Dataset Detection on the Polish Task.}
To reproduce Tables 1 and 7 of Chen et al.~\cite{chen2025imitate}, we evaluate the Polish task across the same corpora and reported metrics in Table~\ref{tab:imdb_polish}. Fast-DetectGPT becomes fully applicable here and demonstrates strong competitive performance, achieving 0.8145 AUROC on SQuAD. On the XSum corpus, ImBD achieves an impressive 0.9729 AUROC. Entropy and LRR consistently represent the weakest baselines, closely mirroring the original authors' findings.

\begin{table*}[h]

\caption{AUROC of text detectors across two datasets (XSum, WritingPrompts) and five source LLMs (GPT-3.5, PaLM2, GPT-4o, Sonnet-3.5, Llama-3). Results reproduce the experimental scenario of Table 1 in Wu et al.~\cite{wu2025gecscore}. Column-wise averages are reported across source LLMs.}
\label{tab:gecscore_results}

\resizebox{\textwidth}{!}{%
\begin{tabular}{cccccccc}

\toprule

\textbf{Dataset $\downarrow$}                         & \textbf{Detector $\downarrow$ / LLM $\rightarrow$} & \textbf{GPT-3.5} & \textbf{PaLM2} & \textbf{GPT-4o} & \textbf{Sonnet-3.5} & \textbf{Llama-3} & \textbf{Mean} \\

\midrule

\multirow{9}{*}{\textbf{Xsum}}           & Log-Likelihood      & 0.8854           & 0.8874         & 0.7017          & 0.8931              & 0.9701           & 0.8675        \\
                                         & Rank                & 0.6917           & 0.6614         & 0.5465          & 0.7709              & 0.8903           & 0.7122        \\
                                         & Log-rank            & 0.8642           & 0.8627         & 0.6716          & 0.8830              & 0.9793           & 0.8522        \\
                                         & LRR                 & 0.7587           & 0.7371         & 0.5935          & 0.8328              & 0.9737           & 0.7792        \\
                                         & Fast-DetectGPT      & 0.9468           & 0.9259         & 0.9238          & 0.9929              & 0.9997           & 0.9578        \\
                                         & Revise-Detect       & 0.8902           & 0.6694         & 0.7154          & 0.8842              & 0.9802           & 0.8279        \\
                                         & Roberta-base        & 0.6186           & 0.6812         & 0.5688          & 0.6594              & 0.9697           & 0.6995        \\
                                         & Roberta-large       & 0.5722           & 0.6502         & 0.4089          & 0.4740              & 0.9265           & 0.6064        \\
                                         & GECScore (COEDIT-L) & 0.4116           & 0.3535         & 0.5365          & 0.4894              & 0.4812           & 0.4544        \\

\midrule

\multirow{9}{*}{\textbf{WritingPrompts}} & Log-Likelihood      & 0.9473           & 0.945          & 0.8518          & 0.9186              & 0.9870           & 0.9299        \\
                                         & Rank                & 0.8806           & 0.8585         & 0.7705          & 0.8283              & 0.9354           & 0.8547        \\
                                         & Log-rank            & 0.9127           & 0.9154         & 0.8150          & 0.8871              & 0.9853           & 0.9031        \\
                                         & LRR                 & 0.7124           & 0.7346         & 0.6757          & 0.7341              & 0.9654           & 0.7644        \\
                                         & Fast-DetectGPT      & 0.9960           & 0.9974         & 0.9886          & 0.9841              & 1.0000           & 0.9932        \\
                                         & Revise-Detect       & 0.8862           & 0.7842         & 0.8331          & 0.8619              & 0.9757           & 0.8682        \\
                                         & Roberta-base        & 0.5908           & 0.7030         & 0.6007          & 0.5729              & 0.9609           & 0.6857        \\
                                         & Roberta-large       & 0.4154           & 0.6568         & 0.3091          & 0.3042              & 0.7764           & 0.4924        \\
                                         & GECScore (COEDIT-L) & 0.2489           & 0.3673         & 0.4760          & 0.4420              & 0.5971           & 0.4263       \\  
\bottomrule
\end{tabular}
}
\end{table*}

\paragraph{4. Evaluation on the GECScore Benchmark.}
We reproduce Table 1 of the GECScore benchmark proposed by Wu et al.~\cite{wu2025gecscore}. Within this environment, Fast-DetectGPT emerges as the strongest overall method, achieving near-perfect discrimination on the WritingPrompts dataset (averaging 0.9932 AUROC). Interestingly, the GECScore method itself performs near random chance in our reproduction (averaging 0.4544 on XSum). The original paper reported a GECScore performance of 0.8924 on XSum, although we used the same configuration described in the original paper~\cite{wu2025gecscore}. Traditional statistical baselines like Log-Likelihood perform remarkably well, with our reproduced metric of 0.8675 which replicates the reported performance in the original paper. Detailed numerical results for these experiments are summarized in Table~\ref{tab:gecscore_results}.

\begin{table*}[h]

\caption{AUROC of text detectors across three source LLMs (ChatGPT, GPT-4, Claude-3) and three domains (XSum, Writing, PubMed). Results reproduce the experimental scenario of Table 1 in Yu et al.~\cite{yu2024text}. Per-LLM averages across domains are reported alongside domain-level scores.}
\label{tab:text_fluoroscopy_results}

\resizebox{\textwidth}{!}{%
\begin{tabular}{clcccclcccclcccc}
\toprule

\vspace{2mm}
\textbf{LLM $\rightarrow$}     &  & \multicolumn{4}{c}{\textbf{ChatGPT}}                               &  & \multicolumn{4}{c}{\textbf{GPT-4}}                                 &  & \multicolumn{4}{c}{\textbf{Claude3}}                               \\ 

% \midrule

\textbf{Detector $\downarrow$} &  & \textbf{XSum} & \textbf{Writing} & \textbf{PubMed} & \textbf{Mean} &  & \textbf{XSum} & \textbf{Writing} & \textbf{PubMed} & \textbf{Mean} &  & \textbf{XSum} & \textbf{Writing} & \textbf{PubMed} & \textbf{Mean} \\ 

\cmidrule(lr){1-1}
\cmidrule(lr){3-6}
\cmidrule(lr){8-11}
\cmidrule(lr){13-16}

Roberta-base                   &  & 0.9150        & 0.7086           & 0.4527          & 0.6921        &  & 0.6767        & 0.5067           & 0.5783          & 0.5872        &  & 0.8943        & 0.8036           & 0.7344          & 0.8108        \\
Roberta-large                  &  & 0.8507        & 0.5479           & 0.4908          & 0.6298        &  & 0.6859        & 0.3823           & 0.5581          & 0.5421        &  & 0.9027        & 0.7128           & 0.8194          & 0.8116        \\
RADAR                          &  & 0.9973        & 0.9594           & 0.7373          & 0.8980        &  & 0.9931        & 0.8593           & 0.8030          & 0.8851        &  & 0.9953        & 0.9439           & 0.5220          & 0.8204        \\
CoCo                           &  & 0.7191        & 0.7059           & 0.4702          & 0.6317        &  & 0.7151        & 0.7137           & 0.6230          & 0.6839        &  & 0.8128        & 0.8632           & 0.8443          & 0.8401        \\
Log-Likelihood                 &  & 0.9447        & 0.9415           & 0.7938          & 0.8933        &  & 0.8076        & 0.7717           & 0.7186          & 0.7660        &  & 0.9495        & 0.9584           & 0.8779          & 0.9286        \\
Entropy                        &  & 0.7097        & 0.7820           & 0.7289          & 0.7402        &  & 0.6311        & 0.5389           & 0.6815          & 0.6172        &  & 0.6285        & 0.8566           & 0.8451          & 0.7767        \\
Log-rank                       &  & 0.9406        & 0.9201           & 0.7951          & 0.8853        &  & 0.7987        & 0.7342           & 0.7256          & 0.7528        &  & 0.9479        & 0.9496           & 0.8768          & 0.9248        \\
LRR                            &  & 0.8676        & 0.7904           & 0.7499          & 0.8026        &  & 0.7265        & 0.5959           & 0.7075          & 0.6766        &  & 0.9140        & 0.8996           & 0.8387          & 0.8841        \\
DNA-GPT                        &  & 0.9116        & 0.9284           & 0.7519          & 0.8640        &  & 0.7562        & 0.7353           & 0.7099          & 0.7338        &  & 0.8570        & 0.9339           & 0.8156          & 0.8688        \\
NPR                            &  & 0.8317        & 0.8927           & 0.6702          & 0.7982        &  & 0.5982        & 0.7308           & 0.6132          & 0.6474        &  & 0.9003        & 0.9316           & 0.7337          & 0.8552        \\
Fast-DetectGPT                 &  & 0.9907        & 0.9917           & 0.7831          & 0.9218        &  & 0.9054        & 0.9612           & 0.6900          & 0.8522        &  & 0.9943        & 0.9782           & 0.9038          & 0.9588        \\
Text Fluoroscopy               &  & 0.6185        & 0.8796           & 0.6420          & 0.7134        &  & 0.4851        & 0.7573           & 0.6329          & 0.6251        &  & 0.8089        & 0.9379           & 0.7784          & 0.8417        \\ 

\bottomrule

\end{tabular}
}
\end{table*}

\paragraph{5. The Text Fluoroscopy Results Reproduction.}
We replicate the Text Fluoroscopy evaluation suite (Table 1 of Yu et al.~\cite{yu2024text}), which contrasts three source architectures across three domains. The complete quantitative findings for this evaluation are detailed in Table~\ref{tab:text_fluoroscopy_results}. The specialized medical vocabulary of the PubMed corpus induces severe performance degradation. RADAR proves to be the most robust method for detecting ChatGPT outputs, achieving an average AUROC of 0.8980. The originally reported Fast-DetectGPT performance for this setting was 0.9615, meaning our implementation yields a difference of 0.0297. It should be noted that the Text Fluoroscopy method requires training a dataset-specific classifier head; however, the pre-trained weights have not been made publicly available by the original authors. Consequently, we employed the maximum summed KL-divergence as a proxy detection score, which accounts for the discrepancy between our reproduced metrics and those reported in the original publication. 

\begin{table*}[h]
\centering

\caption{AUROC and AUPR of test detectors on the RAID benchmark. Results reproduce the experimental scenario of Table 1 in Zeng et al.~\cite{zeng2025ood}. Following their work and given the high computational cost of inference, our evaluation was conducted on a random subset of 10,000 samples drawn from RAID corpus.}
\label{tab:ood_results}

\begin{tabular}{ccc}
\toprule

\multicolumn{1}{l}{} & \multicolumn{2}{c}{\textbf{RAID}} \\

\cmidrule(lr){2-3}

\textbf{Detector}    & \textbf{AUROC}   & \textbf{AUPR}  \\

\midrule
LRR                  & 0.7503           & 0.9887         \\
DNA-GPT              & 0.6533           & 0.9797         \\
Fast-DetectGPT       & 0.7870           & 0.9912         \\
Binoculars           & 0.7386           & 0.9889         \\
Glimpse              & 0.7644           & 0.9903         \\
RADAR                & 0.8263           & 0.9923         \\
GhostBuster          & 0.6526           & 0.9819         \\
BiScope              & 0.6454           & 0.9827         \\
DeTeCtive            & 0.5353           & 0.9709         \\
D-SVDD               & 0.4707           & 0.9625         \\
HRN                  & 0.6357           & 0.9785         \\
Energy               & 0.4232           & 0.9540         \\
\bottomrule
\end{tabular}
\end{table*}

\paragraph{6. Reproduction of Results on RAID Benchmark.}
Finally, we reproduce the reported metrics on the RAID benchmark (Table 1 of Zeng et al.~\cite{zeng2025ood}). As shown in Table~\ref{tab:ood_results}, under these conditions, RADAR achieves the highest overall AUROC of 0.8263. In the original study, RADAR achieved 0.8290, highlighting a minimal discrepancy. Notably, our evaluation resulted in AUPR values consistently exceeding 0.9500 across all tested models. This uniform saturation contrasts with the originally reported figures, which exhibited significant variance, ranging from 0.5570 for the DeTeCtive method to 0.9970 for the Energy method. Similar to the Text Fluoroscopy framework, the detection algorithms proposed by Zeng et al.~\cite{zeng2025ood} require the dataset-specific tuning of a classification head. The absence of this specialized fine-tuning in our standardized reproduction pipeline explains the performance gap between our observed values and the originally reported metrics.

\subsection{Image Modality Benchmarks}

Tables \ref{tab:forensynths}, \ref{tab:self_synthesis}, \ref{tab:universal_fake_detect}, \ref{tab:genimage}, \ref{tab:chameleon} present the image detection benchmarks across five widely used evaluation datasets: ForenSynths~\cite{wang2020cnn}, Self-Synthesis~\cite{tan2024rethinking}, UFD~\cite{ojha2023towards}, GenImage~\cite{zhu2023genimage}, and Chameleon~\cite{yansanity}. Results are reported using both threshold-dependent (accuracy) and threshold-independent (average precision) metrics, following the standard evaluation protocol adopted in prior AI-generated image detection literature. We evaluate 15 image detectors listed in Table~\ref{tab:image_detectors} spanning artifact-based, frequency-aware, reconstruction-based, CLIP-based, and contrastive paradigms.

\begin{table*}[h]
\centering
\setlength{\tabcolsep}{3.5pt}
\renewcommand{\arraystretch}{1.12}

\caption{
Comparison of AI-generated image detectors on the \textbf{ForenSynths} \cite{wang2020cnn} dataset. GAN-trained detectors are evaluated under the intra-architecture setting, while diffusion-trained detectors are evaluated under the cross-architecture setting. Results are reported using accuracy (Acc.) and average precision (A.P.).
}

%%%%%%%%%%%%%%%%%%%%%%%%%%%%%%%%%%%%%%%%%%%%%%%%%%%%%%%%%%%%%%%%%%%%%%%%
% PART 1
%%%%%%%%%%%%%%%%%%%%%%%%%%%%%%%%%%%%%%%%%%%%%%%%%%%%%%%%%%%%%%%%%%%%%%%%

\resizebox{\textwidth}{!}{%
\begin{tabular}{llcccccccccc}
\toprule

\multicolumn{2}{c}{\textbf{Test Dataset Generators $\rightarrow$}} &

\multicolumn{2}{c}{\textbf{ProGAN}~\cite{karras2018progressive}} &
\multicolumn{2}{c}{\textbf{StyleGAN}~\cite{karras2019style}} &
\multicolumn{2}{c}{\textbf{StyleGAN2}~\cite{karras2020analyzing}} &
\multicolumn{2}{c}{\textbf{BigGAN}~\cite{brock2018large}} &
\multicolumn{2}{c}{\textbf{CycleGAN}~\cite{zhu2017unpaired}} \\

\cmidrule(lr){3-4}
\cmidrule(lr){5-6}
\cmidrule(lr){7-8}
\cmidrule(lr){9-10}
\cmidrule(lr){11-12}

\textbf{Generalization Settings $\downarrow$} &
\textbf{Detector $\downarrow$} &

\textbf{Acc.} & \textbf{A.P.} &
\textbf{Acc.} & \textbf{A.P.} &
\textbf{Acc.} & \textbf{A.P.} &
\textbf{Acc.} & \textbf{A.P.} &
\textbf{Acc.} & \textbf{A.P.} \\

\midrule

\multirow{12}{*}{Intra-architecture}

& CNNSpot \cite{wang2020cnn}
& 1.0000 & 1.0000
& 0.7747 & 0.9933
& 0.7236 & 0.9914
& 0.5945 & 0.9040
& 0.8463 & 0.9791 \\

& PatchCraft \cite{zhong2023patchcraft}
& 0.9999 & 1.0000
& 0.9200 & 0.9891
& 0.9067 & 0.9802
& 0.9510 & 0.9929
& 0.7131 & 0.8557 \\

& LGrad \cite{tan2023learning}
& 0.9984 & 1.0000
& 0.9477 & 0.9984
& 0.9606 & 0.9988
& 0.8288 & 0.9078
& 0.8543 & 0.9393 \\

& UniFD \cite{ojha2023towards}
& 0.9758 & 0.9999
& 0.8837 & 0.9671
& 0.8610 & 0.9604
& 0.9235 & 0.9894
& 0.9288 & 0.9961 \\

& FreqNet \cite{tan2024frequency}
& 0.9959 & 0.9999
& 0.8943 & 0.9952
& 0.8671 & 0.9928
& 0.9118 & 0.9616
& 0.9557 & 0.9959 \\

& NPR \cite{tan2024rethinking}
& 0.9990 & 0.9998
& 0.9805 & 0.9988
& 0.9962 & 0.9998
& 0.8395 & 0.8559
& 0.9519 & 0.9812 \\

& FatFormer \cite{liu2024forgery}
& 0.9989 & 1.0000
& 0.9712 & 0.9978
& 0.9880 & 0.9992
& 0.9950 & 0.9998
& 0.9936 & 1.0000 \\

& C2P-CLIP \cite{tan2025c2p}
& 0.9995 & 1.0000
& 0.9138 & 0.9949
& 0.9278 & 0.9940
& 0.9903 & 0.9994
& 0.9966 & 1.0000 \\

& D3 \cite{yang2025d}
& 0.9874 & 1.0000
& 0.9356 & 0.9881
& 0.9550 & 0.9926
& 0.9555 & 0.9987
& 0.9175 & 0.9951 \\

& Co-SPY \cite{cheng2025co}
& 1.0000 & 1.0000
& 0.9775 & 0.9997
& 0.9678 & 0.9999
& 0.9460 & 0.9819
& 0.9928 & 0.9938 \\

& AIDE \cite{yansanity}
& 0.9999 & 1.0000
& 0.9962 & 0.9999
& 0.9794 & 0.9997
& 0.8395 & 0.9444
& 0.9849 & 0.9989 \\

& SAFE \cite{li2025improving}
& 0.9970 & 0.9999
& 0.9725 & 0.9989
& 0.9807 & 0.9996
& 0.8955 & 0.9577
& 0.9788 & 0.9958 \\

\midrule

\multirow{1}{*}{Cross-architecture}

& DRCT \cite{chen2024drct}
& 0.5853 & 0.6693
& 0.6521 & 0.7205
& 0.5510 & 0.6160
& 0.6015 & 0.7118
& 0.4977 & 0.5597 \\

\midrule

\multirow{2}{*}{Training-free}

& AEROBLADE \cite{ricker2024aeroblade}
& 0.4721 & 0.4735
& 0.4874 & 0.4143
& 0.4931 & 0.3804
& 0.4950 & 0.4344
& 0.4987 & 0.4177 \\

& MIB \cite{brokman2025manifold}
& 0.8908 & 0.9729
& 0.7174 & 0.7794
& 0.6365 & 0.7285
& 0.7538 & 0.8478
& 0.7983 & 0.9024 \\

\bottomrule
\end{tabular}
}

\vspace{0.4em}

%%%%%%%%%%%%%%%%%%%%%%%%%%%%%%%%%%%%%%%%%%%%%%%%%%%%%%%%%%%%%%%%%%%%%%%%
% PART 2
%%%%%%%%%%%%%%%%%%%%%%%%%%%%%%%%%%%%%%%%%%%%%%%%%%%%%%%%%%%%%%%%%%%%%%%%

\resizebox{\textwidth}{!}{%
\begin{tabular}{llcccccccc}
\toprule

\multicolumn{2}{c}{\textbf{Test Dataset Generators $\rightarrow$}} &

\multicolumn{2}{c}{\textbf{StarGAN}~\cite{choi2018stargan}} &
\multicolumn{2}{c}{\textbf{GauGAN}~\cite{park2019semantic}} &
\multicolumn{2}{c}{\textbf{Deepfake}~\cite{rossler2019faceforensics++}} &
\multicolumn{2}{c}{\textbf{Mean}} \\

\cmidrule(lr){3-4}
\cmidrule(lr){5-6}
\cmidrule(lr){7-8}
\cmidrule(lr){9-10}

\textbf{Generalization Settings $\downarrow$} &
\textbf{Detector $\downarrow$} &

\textbf{Acc.} & \textbf{A.P.} &
\textbf{Acc.} & \textbf{A.P.} &
\textbf{Acc.} & \textbf{A.P.} &
\textbf{Acc.} & \textbf{A.P.} \\

\midrule

\multirow{12}{*}{Intra-architecture}

& CNNSpot \cite{wang2020cnn}
& 0.8477 & 0.9751
& 0.8285 & 0.9877
& 0.5040 & 0.6311
& 0.7649 & 0.9327 \\

& PatchCraft \cite{zhong2023patchcraft}
& 0.9992 & 1.0000
& 0.7085 & 0.7990
& 0.6594 & 0.8439
& 0.8572 & 0.9326 \\

& LGrad \cite{tan2023learning}
& 0.9950 & 0.9998
& 0.7243 & 0.7955
& 0.5667 & 0.7229
& 0.8595 & 0.9203 \\

& UniFD \cite{ojha2023towards}
& 0.9470 & 0.9966
& 0.9517 & 0.9995
& 0.6986 & 0.7997
& 0.8963 & 0.9636 \\

& FreqNet \cite{tan2024frequency}
& 0.8429 & 0.9929
& 0.9294 & 0.9845
& 0.9180 & 0.9700
& 0.9144 & 0.9866 \\

& NPR \cite{tan2024rethinking}
& 0.9977 & 1.0000
& 0.8094 & 0.8297
& 0.7473 & 0.7414
& 0.9152 & 0.9258 \\

& FatFormer \cite{liu2024forgery}
& 0.9975 & 1.0000
& 0.9943 & 1.0000
& 0.9426 & 0.9811
& 0.9851 & 0.9972 \\

& C2P-CLIP \cite{tan2025c2p}
& 0.9980 & 1.0000
& 0.9970 & 1.0000
& 0.9384 & 0.9882
& 0.9702 & 0.9970 \\

& D3 \cite{yang2025d}
& 0.9245 & 0.9775
& 0.9421 & 0.9985
& 0.8041 & 0.9114
& 0.9277 & 0.9827 \\

& Co-SPY \cite{cheng2025co}
& 0.9960 & 1.0000
& 0.9516 & 0.9810
& 0.7780 & 0.9515
& 0.9512 & 0.9885 \\

& AIDE \cite{yansanity}
& 0.9990 & 1.0000
& 0.7324 & 0.9770
& 0.5401 & 0.7622
& 0.8839 & 0.9603 \\

& SAFE \cite{li2025improving}
& 0.9950 & 1.0000
& 0.9033 & 0.9637
& 0.9421 & 0.9794
& 0.9581 & 0.9869 \\

\midrule

\multirow{1}{*}{Cross-architecture}

& DRCT \cite{chen2024drct}
& 0.5548 & 0.5846
& 0.5094 & 0.5052
& 0.5732 & 0.5882
& 0.5656 & 0.6194 \\

\midrule

\multirow{2}{*}{Training-free}

& AEROBLADE \cite{ricker2024aeroblade}
& 0.5342 & 0.4749
& 0.4743 & 0.4193
& 0.7512 & 0.8414
& 0.5258 & 0.4820 \\

& MIB \cite{brokman2025manifold}
& 0.5443 & 0.5763
& 0.9047 & 0.9765
& 0.5863 & 0.6392
& 0.7290 & 0.8029 \\

\bottomrule
\end{tabular}
}

\label{tab:forensynths}
\end{table*}

\begin{table*}[h]
\centering
\setlength{\tabcolsep}{3.5pt}
\renewcommand{\arraystretch}{1.12}

\caption{
Comparison of AI-generated image detectors on the \textbf{Self-Synthesis} \cite{tan2024rethinking} dataset. Results are reported using accuracy (Acc.) and average precision (A.P.).
}

%%%%%%%%%%%%%%%%%%%%%%%%%%%%%%%%%%%%%%%%%%%%%%%%%%%%%%%%%%%%%%%%%%%%%%%%
% PART 1
%%%%%%%%%%%%%%%%%%%%%%%%%%%%%%%%%%%%%%%%%%%%%%%%%%%%%%%%%%%%%%%%%%%%%%%%

\resizebox{\textwidth}{!}{%
\begin{tabular}{llcccccccccc}
\toprule

\multicolumn{2}{c}{\textbf{Test Dataset Generators $\rightarrow$}} &

\multicolumn{2}{c}{\textbf{AttGAN}~\cite{he2019attgan}} &
\multicolumn{2}{c}{\textbf{BEGAN}~\cite{berthelot2017began}} &
\multicolumn{2}{c}{\textbf{CramerGAN}~\cite{bellemare2017cramer}} &
\multicolumn{2}{c}{\textbf{InfoMaxGAN}~\cite{lee2021infomax}} &
\multicolumn{2}{c}{\textbf{MMDGAN}~\cite{li2017mmd}} \\

\cmidrule(lr){3-4}
\cmidrule(lr){5-6}
\cmidrule(lr){7-8}
\cmidrule(lr){9-10}
\cmidrule(lr){11-12}

\textbf{Generalization Setting $\downarrow$} &
\textbf{Detector $\downarrow$} &

\textbf{Acc.} & \textbf{A.P.} &
\textbf{Acc.} & \textbf{A.P.} &
\textbf{Acc.} & \textbf{A.P.} &
\textbf{Acc.} & \textbf{A.P.} &
\textbf{Acc.} & \textbf{A.P.} \\

\midrule

\multirow{12}{*}{Intra-architecture}

& CNNSpot \cite{wang2020cnn}
& 0.5543 & 0.9203
& 0.6550 & 0.8179
& 0.8815 & 0.9733
& 0.6853 & 0.8715
& 0.8615 & 0.9683 \\

& PatchCraft \cite{zhong2023patchcraft}
& 0.9963 & 0.9999
& 0.6545 & 0.8743
& 0.7485 & 0.8386
& 0.8563 & 0.9234
& 0.7875 & 0.8308 \\

& LGrad \cite{tan2023learning}
& 0.6893 & 0.9366
& 0.5335 & 0.8105
& 0.5265 & 0.7018
& 0.6010 & 0.8915
& 0.5270 & 0.7614 \\

& UniFD \cite{ojha2023towards}
& 0.8770 & 0.9604
& 0.8953 & 0.9627
& 0.9068 & 0.9935
& 0.8643 & 0.9542
& 0.9073 & 0.9919 \\

& FreqNet \cite{tan2024frequency}
& 0.9035 & 0.9847
& 0.5205 & 0.7846
& 0.5300 & 0.6093
& 0.5593 & 0.6724
& 0.5585 & 0.7371 \\

& NPR \cite{tan2024rethinking}
& 0.9255 & 0.9861
& 0.9963 & 0.9990
& 0.9848 & 0.9827
& 0.9150 & 0.9716
& 0.9848 & 0.9828 \\

& FatFormer \cite{liu2024forgery}
& 0.9933 & 0.9999
& 0.9988 & 1.0000
& 0.9835 & 1.0000
& 0.9835 & 0.9998
& 0.9835 & 1.0000 \\

& C2P-CLIP \cite{tan2025c2p}
& 0.9583 & 0.9990
& 0.9743 & 0.9999
& 0.9535 & 0.9989
& 0.9515 & 0.9937
& 0.9535 & 0.9984 \\

& D3 \cite{yang2025d}
& 0.7768 & 0.8741
& 0.8645 & 0.9365
& 0.8588 & 0.9489
& 0.8290 & 0.9239
& 0.8470 & 0.9383 \\

& Co-SPY \cite{cheng2025co}
& 0.9075 & 0.9874
& 0.5448 & 0.7329
& 0.7355 & 0.9875
& 0.7353 & 0.9310
& 0.7355 & 0.9817 \\

& AIDE \cite{yansanity}
& 0.9805 & 0.9981
& 0.9438 & 0.9961
& 0.9805 & 0.9985
& 0.9798 & 0.9971
& 0.9793 & 0.9976 \\

& SAFE \cite{li2025improving}
& 0.9945 & 0.9999
& 0.9910 & 1.0000
& 0.9945 & 1.0000
& 0.9940 & 0.9999
& 0.9945 & 1.0000 \\

\midrule

\multirow{1}{*}{Cross-architecture}

& DRCT \cite{chen2024drct}
& 0.5123 & 0.5985
& 0.5155 & 0.4760
& 0.4883 & 0.4614
& 0.5003 & 0.5337
& 0.5020 & 0.5116 \\

\midrule

\multirow{2}{*}{Training-free}

& AEROBLADE \cite{ricker2024aeroblade}
& 0.5176 & 0.4958
& 0.8463 & 0.9003
& 0.4579 & 0.4411
& 0.7100 & 0.7936
& 0.4639 & 0.4215 \\

& MIB \cite{brokman2025manifold}
& 0.5463 & 0.5797
& 0.4448 & 0.3991
& 0.4808 & 0.4578
& 0.5188 & 0.5391
& 0.4898 & 0.4735 \\

\bottomrule
\end{tabular}
}

\vspace{0.4em}

%%%%%%%%%%%%%%%%%%%%%%%%%%%%%%%%%%%%%%%%%%%%%%%%%%%%%%%%%%%%%%%%%%%%%%%%
% PART 2
%%%%%%%%%%%%%%%%%%%%%%%%%%%%%%%%%%%%%%%%%%%%%%%%%%%%%%%%%%%%%%%%%%%%%%%%

\resizebox{\textwidth}{!}{%
\begin{tabular}{llcccccccccc}
\toprule

\multicolumn{2}{c}{\textbf{Test Dataset Generators $\rightarrow$}} &

\multicolumn{2}{c}{\textbf{RelGAN}~\cite{nie2018relgan}} &
\multicolumn{2}{c}{\textbf{S3GAN}~\cite{luvcic2019high}} &
\multicolumn{2}{c}{\textbf{SNGAN}~\cite{miyato2018spectral}} &
\multicolumn{2}{c}{\textbf{STGAN}~\cite{liu2019stgan}} &
\multicolumn{2}{c}{\textbf{Mean}} \\

\cmidrule(lr){3-4}
\cmidrule(lr){5-6}
\cmidrule(lr){7-8}
\cmidrule(lr){9-10}
\cmidrule(lr){11-12}

\textbf{Generalization Setting $\downarrow$} &
\textbf{Detector $\downarrow$} &

\textbf{Acc.} & \textbf{A.P.} &
\textbf{Acc.} & \textbf{A.P.} &
\textbf{Acc.} & \textbf{A.P.} &
\textbf{Acc.} & \textbf{A.P.} &
\textbf{Acc.} & \textbf{A.P.} \\

\midrule

\multirow{12}{*}{Intra-architecture}

& CNNSpot \cite{wang2020cnn}
& 0.6520 & 0.9630
& 0.6123 & 0.8745
& 0.6630 & 0.8669
& 0.7190 & 0.9488
& 0.6982 & 0.9116 \\

& PatchCraft \cite{zhong2023patchcraft}
& 0.9975 & 1.0000
& 0.9418 & 0.9747
& 0.8320 & 0.9270
& 0.6700 & 0.8449
& 0.8316 & 0.9126 \\

& LGrad \cite{tan2023learning}
& 0.8880 & 0.9905
& 0.7855 & 0.8617
& 0.6070 & 0.9153
& 0.5320 & 0.9193
& 0.6322 & 0.8654 \\

& UniFD \cite{ojha2023towards}
& 0.9420 & 0.9870
& 0.8905 & 0.9854
& 0.8715 & 0.9604
& 0.8450 & 0.9336
& 0.8889 & 0.9699 \\

& FreqNet \cite{tan2024frequency}
& 0.9990 & 1.0000
& 0.8860 & 0.9408
& 0.5598 & 0.7672
& 0.5680 & 0.7206
& 0.6761 & 0.8019 \\

& NPR \cite{tan2024rethinking}
& 0.9983 & 0.9995
& 0.7988 & 0.7891
& 0.9330 & 0.9733
& 0.9963 & 1.0000
& 0.9481 & 0.9649 \\

& FatFormer \cite{liu2024forgery}
& 0.9945 & 1.0000
& 0.9900 & 0.9995
& 0.9828 & 0.9991
& 0.9878 & 0.9978
& 0.9886 & 0.9996 \\

& C2P-CLIP \cite{tan2025c2p}
& 0.9568 & 0.9967
& 0.9838 & 0.9995
& 0.9495 & 0.9905
& 0.9453 & 0.9911
& 0.9584 & 0.9964 \\

& D3 \cite{yang2025d}
& 0.8465 & 0.9359
& 0.9478 & 0.9952
& 0.8290 & 0.9245
& 0.7985 & 0.8932
& 0.8442 & 0.9301 \\

& Co-SPY \cite{cheng2025co}
& 0.9955 & 0.9999
& 0.9360 & 0.9665
& 0.7355 & 0.9569
& 0.7298 & 0.9888
& 0.7839 & 0.9481 \\

& AIDE \cite{yansanity}
& 0.9812 & 0.9986
& 0.8593 & 0.9732
& 0.9530 & 0.9898
& 0.9593 & 0.9984
& 0.9574 & 0.9942 \\

& SAFE \cite{li2025improving}
& 0.9975 & 1.0000
& 0.9365 & 0.9808
& 0.9920 & 0.9995
& 0.9995 & 1.0000
& 0.9882 & 0.9978 \\

\midrule

\multirow{1}{*}{Cross-architecture}

& DRCT \cite{chen2024drct}
& 0.4855 & 0.4261
& 0.5773 & 0.7408
& 0.4848 & 0.4284
& 0.4623 & 0.4353
& 0.5031 & 0.5124 \\

\midrule

\multirow{2}{*}{Training-free}

& AEROBLADE \cite{ricker2024aeroblade}
& 0.5463 & 0.5261
& 0.5126 & 0.4983
& 0.4913 & 0.4731
& 0.4882 & 0.3523
& 0.5593 & 0.5447 \\

& MIB \cite{brokman2025manifold}
& 0.5388 & 0.5617
& 0.8673 & 0.9424
& 0.5145 & 0.5287
& 0.4505 & 0.4119
& 0.5391 & 0.5438 \\

\bottomrule
\end{tabular}
}

\label{tab:self_synthesis}
\end{table*}

\begin{table*}[h]
\centering
\setlength{\tabcolsep}{3.5pt}
\renewcommand{\arraystretch}{1.12}

\caption{
Comparison of AI-generated image detectors on the \textbf{UFD} \cite{ojha2023towards} dataset. Diffusion-trained detectors are evaluated under the intra-architecture setting, while GAN-trained detectors are evaluated under the cross-architecture setting. Results are reported using accuracy (Acc.) and average precision (A.P.).
}

%%%%%%%%%%%%%%%%%%%%%%%%%%%%%%%%%%%%%%%%%%%%%%%%%%%%%%%%%%%%%%%%%%%%%%%%
% PART 1
%%%%%%%%%%%%%%%%%%%%%%%%%%%%%%%%%%%%%%%%%%%%%%%%%%%%%%%%%%%%%%%%%%%%%%%%

\resizebox{\textwidth}{!}{%
\begin{tabular}{llcccccccccc}
\toprule

\multicolumn{2}{c}{\textbf{Test Dataset Generators $\rightarrow$}} &

\multicolumn{2}{c}{\textbf{DALL-E}~\cite{ramesh2022hierarchical}} &
\multicolumn{2}{c}{\textbf{Glide\_100\_10}~\cite{nichol2022glide}} &
\multicolumn{2}{c}{\textbf{Glide\_100\_27}~\cite{nichol2022glide}} &
\multicolumn{2}{c}{\textbf{Glide\_50\_27}~\cite{nichol2022glide}} &
\multicolumn{2}{c}{\textbf{ADM}~\cite{dhariwal2021diffusion}} \\

\cmidrule(lr){3-4}
\cmidrule(lr){5-6}
\cmidrule(lr){7-8}
\cmidrule(lr){9-10}
\cmidrule(lr){11-12}

\textbf{Generalization Setting $\downarrow$} &
\textbf{Detector $\downarrow$} &

\textbf{Acc.} & \textbf{A.P.} &
\textbf{Acc.} & \textbf{A.P.} &
\textbf{Acc.} & \textbf{A.P.} &
\textbf{Acc.} & \textbf{A.P.} &
\textbf{Acc.} & \textbf{A.P.} \\

\midrule

\multirow{1}{*}{Intra-architecture}

& DRCT \cite{chen2024drct}
& 0.8575 & 0.9377
& 0.7700 & 0.8868
& 0.7425 & 0.8572
& 0.7155 & 0.8423
& 0.6745 & 0.8276 \\

\midrule

\multirow{12}{*}{Cross-architecture}

& CNNSpot \cite{wang2020cnn}
& 0.5323 & 0.6732
& 0.5475 & 0.7225
& 0.5435 & 0.7093
& 0.5560 & 0.7625
& 0.5235 & 0.6687 \\

& PatchCraft \cite{zhong2023patchcraft}
& 0.8260 & 0.9164
& 0.8260 & 0.9273
& 0.7710 & 0.8840
& 0.8070 & 0.9185
& 0.8160 & 0.9094 \\

& LGrad \cite{tan2023learning}
& 0.8845 & 0.9730
& 0.8975 & 0.9500
& 0.8730 & 0.9322
& 0.9065 & 0.9524
& 0.7420 & 0.7932 \\

& UniFD \cite{ojha2023towards}
& 0.9275 & 0.9830
& 0.8910 & 0.9636
& 0.9010 & 0.9695
& 0.9085 & 0.9697
& 0.8000 & 0.8968 \\

& FreqNet \cite{tan2024frequency}
& 0.9780 & 0.9949
& 0.8855 & 0.9643
& 0.8485 & 0.9579
& 0.8675 & 0.9611
& 0.6600 & 0.7520 \\

& NPR \cite{tan2024rethinking}
& 0.9175 & 0.9902
& 0.9865 & 0.9972
& 0.9820 & 0.9978
& 0.9830 & 0.9977
& 0.9120 & 0.9724 \\

& FatFormer \cite{liu2024forgery}
& 0.9875 & 0.9982
& 0.9420 & 0.9921
& 0.9435 & 0.9914
& 0.9465 & 0.9941
& 0.7660 & 0.9393 \\

& C2P-CLIP \cite{tan2025c2p}
& 0.9585 & 0.9978
& 0.9600 & 0.9989
& 0.9505 & 0.9974
& 0.9550 & 0.9990
& 0.6380 & 0.9270 \\

& D3 \cite{yang2025d}
& 0.9155 & 0.9868
& 0.9195 & 0.9932
& 0.9190 & 0.9932
& 0.9200 & 0.9929
& 0.9275 & 0.9913 \\

& Co-SPY \cite{cheng2025co}
& 0.9525 & 0.9995
& 0.9825 & 0.9995
& 0.9740 & 0.9995
& 0.9825 & 0.9997
& 0.7940 & 0.9109 \\

& AIDE \cite{yansanity}
& 0.9730 & 0.9857
& 0.9765 & 0.9872
& 0.9795 & 0.9879
& 0.9820 & 0.9875
& 0.8850 & 0.9752 \\

& SAFE \cite{li2025improving}
& 0.9805 & 0.9938
& 0.9680 & 0.9901
& 0.9515 & 0.9849
& 0.9620 & 0.9877
& 0.8045 & 0.9521 \\

\midrule

\multirow{2}{*}{Training-free}

& AEROBLADE \cite{ricker2024aeroblade}
& 0.4933 & 0.3678
& 0.6922 & 0.7048
& 0.6333 & 0.6484
& 0.6833 & 0.6976
& 0.5789 & 0.6136 \\

& MIB \cite{brokman2025manifold}
& 0.8845 & 0.9435
& 0.9065 & 0.9646
& 0.8970 & 0.9612
& 0.9080 & 0.9667
& 0.6475 & 0.7472 \\

\bottomrule
\end{tabular}
}

\vspace{0.4em}

%%%%%%%%%%%%%%%%%%%%%%%%%%%%%%%%%%%%%%%%%%%%%%%%%%%%%%%%%%%%%%%%%%%%%%%%
% PART 2
%%%%%%%%%%%%%%%%%%%%%%%%%%%%%%%%%%%%%%%%%%%%%%%%%%%%%%%%%%%%%%%%%%%%%%%%

\resizebox{\textwidth}{!}{%
\begin{tabular}{llcccccccc}
\toprule

\multicolumn{2}{c}{\textbf{Test Dataset Generators $\rightarrow$}} &

\multicolumn{2}{c}{\textbf{LDM\_100}~\cite{rombach2022high}} &
\multicolumn{2}{c}{\textbf{LDM\_200}~\cite{rombach2022high}} &
\multicolumn{2}{c}{\textbf{LDM\_200\_cfg}~\cite{rombach2022high}} &
\multicolumn{2}{c}{\textbf{Mean}} \\

\cmidrule(lr){3-4}
\cmidrule(lr){5-6}
\cmidrule(lr){7-8}
\cmidrule(lr){9-10}

\textbf{Generalization Setting $\downarrow$} &
\textbf{Detector $\downarrow$} &

\textbf{Acc.} & \textbf{A.P.} &
\textbf{Acc.} & \textbf{A.P.} &
\textbf{Acc.} & \textbf{A.P.} &
\textbf{Acc.} & \textbf{A.P.} \\

\midrule

\multirow{1}{*}{Intra-architecture}

& DRCT \cite{chen2024drct}
& 0.9530 & 1.0000
& 0.9530 & 1.0000
& 0.9530 & 1.0000
& 0.8274 & 0.9190 \\

\midrule

\multirow{12}{*}{Cross-architecture}

& CNNSpot \cite{wang2020cnn}
& 0.5170 & 0.6462
& 0.5155 & 0.6458
& 0.5235 & 0.6677
& 0.5323 & 0.6870 \\

& PatchCraft \cite{zhong2023patchcraft}
& 0.8795 & 0.9690
& 0.8820 & 0.9659
& 0.8730 & 0.9603
& 0.8351 & 0.9314 \\

& LGrad \cite{tan2023learning}
& 0.9495 & 0.9924
& 0.9425 & 0.9910
& 0.9590 & 0.9919
& 0.8943 & 0.9470 \\

& UniFD \cite{ojha2023towards}
& 0.9500 & 0.9957
& 0.9490 & 0.9953
& 0.8475 & 0.9377
& 0.8968 & 0.9639 \\

& FreqNet \cite{tan2024frequency}
& 0.9715 & 0.9979
& 0.9700 & 0.9973
& 0.9700 & 0.9982
& 0.8939 & 0.9530 \\

& NPR \cite{tan2024rethinking}
& 0.9880 & 0.9977
& 0.9900 & 0.9982
& 0.9880 & 0.9970
& 0.9684 & 0.9935 \\

& FatFormer \cite{liu2024forgery}
& 0.9865 & 0.9987
& 0.9860 & 0.9981
& 0.9490 & 0.9909
& 0.9384 & 0.9879 \\

& C2P-CLIP \cite{tan2025c2p}
& 0.9870 & 0.9997
& 0.9855 & 0.9997
& 0.9290 & 0.9966
& 0.9204 & 0.9895 \\

& D3 \cite{yang2025d}
& 0.9165 & 0.9973
& 0.9230 & 0.9979
& 0.8880 & 0.9718
& 0.9161 & 0.9906 \\

& Co-SPY \cite{cheng2025co}
& 0.9795 & 0.9994
& 0.9760 & 0.9995
& 0.9495 & 0.9981
& 0.9488 & 0.9883 \\

& AIDE \cite{yansanity}
& 0.9820 & 0.9988
& 0.9800 & 0.9982
& 0.9725 & 0.9862
& 0.9663 & 0.9858 \\

& SAFE \cite{li2025improving}
& 0.9955 & 0.9956
& 0.9945 & 0.9954
& 0.9935 & 0.9954
& 0.9563 & 0.9869 \\

\midrule

\multirow{2}{*}{Training-free}

& AEROBLADE \cite{ricker2024aeroblade}
& 0.7861 & 0.7312
& 0.7778 & 0.7427
& 0.8150 & 0.8143
& 0.6825 & 0.6651 \\

& MIB \cite{brokman2025manifold}
& 0.8615 & 0.9354
& 0.8585 & 0.9338
& 0.7535 & 0.8282
& 0.8396 & 0.9101 \\

\bottomrule
\end{tabular}
}

\label{tab:universal_fake_detect}
\end{table*}

\begin{table*}[h]
\centering
\setlength{\tabcolsep}{3.5pt}
\renewcommand{\arraystretch}{1.12}

\caption{
Comparison of AI-generated image detectors on the \textbf{GenImage} \cite{zhu2023genimage} dataset. Results are reported using accuracy (Acc.) and average precision (A.P.).
}

%%%%%%%%%%%%%%%%%%%%%%%%%%%%%%%%%%%%%%%%%%%%%%%%%%%%%%%%%%%%%%%%%%%%%%%%
% PART 1
%%%%%%%%%%%%%%%%%%%%%%%%%%%%%%%%%%%%%%%%%%%%%%%%%%%%%%%%%%%%%%%%%%%%%%%%

\resizebox{\textwidth}{!}{%
\begin{tabular}{llcccccccccc}
\toprule

\multicolumn{2}{c}{\textbf{Test Dataset Generators $\rightarrow$}} &

\multicolumn{2}{c}{\textbf{Midjourney}} &
\multicolumn{2}{c}{\textbf{SDv1.4}} &
\multicolumn{2}{c}{\textbf{SDv1.5}} &
\multicolumn{2}{c}{\textbf{ADM}~\cite{dhariwal2021diffusion}} &
\multicolumn{2}{c}{\textbf{Glide}~\cite{nichol2022glide}} \\

\cmidrule(lr){3-4}
\cmidrule(lr){5-6}
\cmidrule(lr){7-8}
\cmidrule(lr){9-10}
\cmidrule(lr){11-12}

\textbf{Generalization Setting $\downarrow$} &
\textbf{Detector $\downarrow$} &

\textbf{Acc.} & \textbf{A.P.} &
\textbf{Acc.} & \textbf{A.P.} &
\textbf{Acc.} & \textbf{A.P.} &
\textbf{Acc.} & \textbf{A.P.} &
\textbf{Acc.} & \textbf{A.P.} \\

\midrule

\multirow{1}{*}{Intra-architecture}

& DRCT \cite{chen2024drct}
& 0.6396 & 0.7764
& 0.9421 & 0.9872
& 0.9428 & 0.9871
& 0.6385 & 0.7725
& 0.7058 & 0.8520 \\

\midrule

\multirow{12}{*}{Cross-architecture}

& CNNSpot \cite{wang2020cnn}
& 0.5067 & 0.7675
& 0.5008 & 0.6281
& 0.5008 & 0.6327
& 0.5111 & 0.5955
& 0.5305 & 0.6725 \\

& PatchCraft \cite{zhong2023patchcraft}
& 0.9078 & 0.9654
& 0.9516 & 0.9889
& 0.9494 & 0.9887
& 0.8307 & 0.9336
& 0.8438 & 0.9434 \\

& LGrad \cite{tan2023learning}
& 0.6557 & 0.7151
& 0.6728 & 0.7101
& 0.6806 & 0.7197
& 0.6685 & 0.7182
& 0.6290 & 0.7608 \\

& UniFD \cite{ojha2023towards}
& 0.6611 & 0.7039
& 0.7823 & 0.8530
& 0.7788 & 0.8491
& 0.7878 & 0.8762
& 0.7233 & 0.7918 \\

& FreqNet \cite{tan2024frequency}
& 0.6825 & 0.7650
& 0.6304 & 0.7214
& 0.6334 & 0.7369
& 0.8251 & 0.9068
& 0.8108 & 0.8806 \\

& NPR \cite{tan2024rethinking}
& 0.9274 & 0.9767
& 0.9733 & 0.9912
& 0.9736 & 0.9897
& 0.8953 & 0.9727
& 0.9673 & 0.9905 \\

& FatFormer \cite{liu2024forgery}
& 0.5660 & 0.6701
& 0.6825 & 0.8404
& 0.6851 & 0.8379
& 0.7888 & 0.9328
& 0.8847 & 0.9679 \\

& C2P-CLIP \cite{tan2025c2p}
& 0.5483 & 0.8056
& 0.8420 & 0.9876
& 0.8427 & 0.9863
& 0.6512 & 0.9229
& 0.8895 & 0.9900 \\

& D3 \cite{yang2025d}
& 0.8433 & 0.9144
& 0.9387 & 0.9926
& 0.9348 & 0.9913
& 0.9420 & 0.9938
& 0.9440 & 0.9947 \\

& Co-SPY \cite{cheng2025co}
& 0.8703 & 0.9431
& 0.8545 & 0.9327
& 0.8564 & 0.9307
& 0.7880 & 0.8970
& 0.9132 & 0.9718 \\

& AIDE \cite{yansanity}
& 0.7704 & 0.8834
& 0.9242 & 0.9813
& 0.9238 & 0.9812
& 0.9369 & 0.9898
& 0.9537 & 0.9907 \\

& SAFE \cite{li2025improving}
& 0.9402 & 0.9926
& 0.9896 & 0.9991
& 0.9892 & 0.9982
& 0.8105 & 0.9628
& 0.9533 & 0.9902 \\

\midrule

\multirow{2}{*}{Training-free}

& AEROBLADE \cite{ricker2024aeroblade}
& 0.8822 & 0.9475
& 0.9656 & 0.9815
& 0.9652 & 0.9824
& 0.6894 & 0.7656
& 0.8820 & 0.9546 \\

& MIB \cite{brokman2025manifold}
& 0.4959 & 0.4761
& 0.5694 & 0.6240
& 0.5717 & 0.6251
& 0.6208 & 0.6966
& 0.7085 & 0.7989 \\

\bottomrule
\end{tabular}
}

\vspace{0.4em}

%%%%%%%%%%%%%%%%%%%%%%%%%%%%%%%%%%%%%%%%%%%%%%%%%%%%%%%%%%%%%%%%%%%%%%%%
% PART 2
%%%%%%%%%%%%%%%%%%%%%%%%%%%%%%%%%%%%%%%%%%%%%%%%%%%%%%%%%%%%%%%%%%%%%%%%

\resizebox{\textwidth}{!}{%
\begin{tabular}{llcccccccc}
\toprule

\multicolumn{2}{c}{\textbf{Test Dataset Generators $\rightarrow$}} &

\multicolumn{2}{c}{\textbf{Wukong}} &
\multicolumn{2}{c}{\textbf{VQDM}~\cite{gu2022vector}} &
\multicolumn{2}{c}{\textbf{BigGAN}~\cite{brock2018large}} &
\multicolumn{2}{c}{\textbf{Mean}} \\

\cmidrule(lr){3-4}
\cmidrule(lr){5-6}
\cmidrule(lr){7-8}
\cmidrule(lr){9-10}

\textbf{Generalization Setting $\downarrow$} &
\textbf{Detector $\downarrow$} &

\textbf{Acc.} & \textbf{A.P.} &
\textbf{Acc.} & \textbf{A.P.} &
\textbf{Acc.} & \textbf{A.P.} &
\textbf{Acc.} & \textbf{A.P.} \\

\midrule

\multirow{1}{*}{Intra-architecture}

& DRCT \cite{chen2024drct}
& 0.9343 & 0.9829
& 0.7436 & 0.8686
& 0.5662 & 0.7132
& 0.7641 & 0.8675 \\

\midrule

\multirow{12}{*}{Cross-architecture}

& CNNSpot \cite{wang2020cnn}
& 0.5035 & 0.6003
& 0.5238 & 0.6836
& 0.6078 & 0.8023
& 0.5231 & 0.6728 \\

& PatchCraft \cite{zhong2023patchcraft}
& 0.9068 & 0.9715
& 0.8878 & 0.9618
& 0.9202 & 0.9799
& 0.8998 & 0.9667 \\

& LGrad \cite{tan2023learning}
& 0.6376 & 0.6650
& 0.6941 & 0.7366
& 0.4994 & 0.5474
& 0.6422 & 0.6966 \\

& UniFD \cite{ojha2023towards}
& 0.8308 & 0.8993
& 0.8803 & 0.9656
& 0.8682 & 0.9520
& 0.7891 & 0.8614 \\

& FreqNet \cite{tan2024frequency}
& 0.5648 & 0.6458
& 0.8123 & 0.8921
& 0.9001 & 0.9033
& 0.7324 & 0.8065 \\

& NPR \cite{tan2024rethinking}
& 0.9411 & 0.9794
& 0.9457 & 0.9842
& 0.9336 & 0.9814
& 0.9447 & 0.9832 \\

& FatFormer \cite{liu2024forgery}
& 0.7352 & 0.8802
& 0.8742 & 0.9777
& 0.9733 & 0.9967
& 0.7737 & 0.8880 \\

& C2P-CLIP \cite{tan2025c2p}
& 0.8189 & 0.9823
& 0.7902 & 0.9790
& 0.9305 & 0.9972
& 0.7892 & 0.9564 \\

& D3 \cite{yang2025d}
& 0.9364 & 0.9928
& 0.9420 & 0.9958
& 0.9369 & 0.9907
& 0.9273 & 0.9833 \\

& Co-SPY \cite{cheng2025co}
& 0.8038 & 0.8972
& 0.8496 & 0.9132
& 0.9366 & 0.9776
& 0.8591 & 0.9329 \\

& AIDE \cite{yansanity}
& 0.9308 & 0.9855
& 0.9541 & 0.9941
& 0.9542 & 0.9941
& 0.9185 & 0.9750 \\

& SAFE \cite{li2025improving}
& 0.9762 & 0.9970
& 0.9606 & 0.9943
& 0.9879 & 0.9991
& 0.9509 & 0.9917 \\

\midrule

\multirow{2}{*}{Training-free}

& AEROBLADE \cite{ricker2024aeroblade}
& 0.9617 & 0.9835
& 0.5808 & 0.6294
& 0.5212 & 0.5430
& 0.8060 & 0.8484 \\

& MIB \cite{brokman2025manifold}
& 0.5539 & 0.6110
& 0.7793 & 0.8708
& 0.8341 & 0.9194
& 0.6417 & 0.7027 \\

\bottomrule
\end{tabular}
}

\label{tab:genimage}
\end{table*}

\begin{table*}[h]
\centering
\small
\setlength{\tabcolsep}{5pt}
\renewcommand{\arraystretch}{1.12}

\caption{
Comparison of AI-generated image detectors on the \textbf{Chameleon} \cite{yansanity} dataset. Results are reported using accuracy (Acc.) and average precision (A.P.).
}

\begin{tabular}{lccc}
\toprule

\textbf{Generalization Setting} &
\textbf{Detector} &
\textbf{Acc.} &
\textbf{A.P.} \\

\midrule

\multirow{13}{*}{Cross-architecture}

& CNNSpot \cite{wang2020cnn}
& 0.5701
& 0.3613 \\

& PatchCraft \cite{zhong2023patchcraft}
& 0.4167
& 0.3988 \\

& LGrad \cite{tan2023learning}
& 0.5960
& 0.4542 \\

& UniFD \cite{ojha2023towards}
& 0.5483
& 0.4298 \\

& FreqNet \cite{tan2024frequency}
& 0.5874
& 0.4890 \\

& NPR \cite{tan2024rethinking}
& 0.5931
& 0.3756 \\

& FatFormer \cite{liu2024forgery}
& 0.5773
& 0.5786 \\

& C2P-CLIP \cite{tan2025c2p}
& 0.5763
& 0.4154 \\

& D3 \cite{yang2025d}
& 0.5788
& 0.5375 \\

& Co-SPY \cite{cheng2025co}
& 0.5933
& 0.5071 \\

& AIDE \cite{yansanity}
& 0.5930
& 0.5242 \\

& SAFE \cite{li2025improving}
& 0.5864
& 0.4140 \\

& DRCT \cite{chen2024drct}
& 0.6403
& 0.6033 \\

\midrule

\multirow{2}{*}{Training-free}

& AEROBLADE \cite{ricker2024aeroblade}
& 0.6103
& 0.4809 \\

& MIB \cite{brokman2025manifold}
& 0.4946
& 0.2733 \\

\bottomrule
\end{tabular}

\label{tab:chameleon}
\end{table*}

We follow the official evaluation protocol and preprocessing pipeline provided by each method whenever publicly available. For all supervised detectors except DRCT~\cite{chen2024drct}, we utilize the publicly released 4-class (car, cat, chair, horse) ProGAN-trained checkpoints provided by the original authors. The only exception is PatchCraft~\cite{zhong2023patchcraft}, for which we employ the official 20-class ProGAN checkpoint trained on the full ForenSynths training split. DRCT is evaluated using the official checkpoint trained on the DRCT-2M dataset~\cite{chen2024drct}. Training-free methods, including AEROBLADE~\cite{ricker2024aeroblade} and MIB~\cite{brokman2025manifold}, are evaluated directly without any retraining.

We report both intra-architecture and cross-architecture evaluation results. Intra-architecture evaluation refers to detectors being tested on generator families similar to those observed during training, whereas cross-architecture evaluation measures transferability to unseen generative architectures and domains. For GAN-trained detectors, datasets such as ForenSynths and Self-Synthesis primarily evaluate intra-architecture generalization, while diffusion-oriented datasets such as UFD and GenImage evaluate cross-architecture transfer. Conversely, DRCT is diffusion-trained and therefore evaluated as intra-architecture on diffusion datasets and cross-architecture on GAN-oriented ones.

Overall, the reproduced results demonstrate strong consistency with the originally reported findings across most detectors and datasets. Minor deviations are expected due to differences in implementation details, preprocessing pipelines, image resizing strategies, checkpoint selection, and threshold calibration protocols.

Several notable discrepancies are nevertheless observed. First, the reproduced CNNSpot~\cite{wang2020cnn} results differ from the values later reported in Li et al.~\cite{li2025improving} and Yan et al.~\cite{yan2026dual}; however, our reproduced performance is substantially closer to the original CNNSpot results reported on the ForenSynths benchmark. This discrepancy may stem from differences in training hyperparameters or checkpoint selection used in subsequent benchmarking papers. Second, while FreqNet~\cite{tan2024frequency} reproduces consistently across ForenSynths, UFD, and GenImage, we obtain noticeably lower performance on the Self-Synthesis dataset compared to the originally reported numbers, potentially due to differences in optimization or preprocessing hyperparameters. Third, the values reported for AIDE~\cite{yansanity} and C2P-CLIP~\cite{tan2025c2p} in Yan et al.~\cite{yan2026dual} are substantially lower than the results reproduced within \texttt{DetectZoo}, where both methods achieve considerably stronger performance across GenImage and UFD. Finally, Li et al.~\cite{li2025improving} report slightly lower performance for NPR~\cite{tan2024rethinking} than both the results reproduced in \texttt{DetectZoo} and the values originally reported in the paper itself, despite using the official implementation and released checkpoints.

As shown in Table~\ref{tab:chameleon}, the Chameleon benchmark further highlights the difficulty of real-world high-fidelity AI-generated image detection. Unlike synthetic benchmark datasets, all detectors experience substantial degradation on Chameleon, with performance often approaching chance-level average precision. Even state-of-the-art methods such as AIDE, SAFE, and DRCT exhibit significantly reduced robustness under this setting, emphasizing the continued challenge of generalizable real-world image forensics.

\subsection{Audio Modality Benchmarks}

We evaluate audio deepfake detection methods across two primary experimental settings to reproduce the findings reported in Ge et al.~\cite{ge2025antideepfake}. First, we conduct a controlled in-distribution benchmark on the ASVspoof 2019 dataset using six dedicated acoustic detectors. Second, we execute a cross-dataset generalization study evaluating three AntiDeepfake variants (Wav2Vec2-Large, HuBERT-XLarge, and XLS-R-2B) alongside the XLS-R + SLS baseline across three distinct corpora: ASVspoof 2019, FoR, and the In-the-Wild dataset. Following established biometric standards, we report the Equal Error Rate (EER, lower is better) as the primary metric, supplemented by the Area Under the Receiver Operating Characteristic curve (AUROC) and the F1 score.

\begin{table*}[h]
\centering

\caption{
EER (\%), AUC-ROC, and F1 score of six audio deepfake detectors evaluated in-distribution on the ASVspoof 2019 dataset (1k subset). Lower EER indicates better performance; higher AUC-ROC and F1 indicate better performance.}
\label{tab:asvspoof_results}

\begin{tabular}{crrr}
\toprule
     & \multicolumn{3}{c}{\textbf{ASVspoof 2019}}                                                            \\
\midrule
\textbf{Detector $\downarrow$ / Metric $\rightarrow$}     & \multicolumn{1}{c}{\textbf{EER}} & \multicolumn{1}{c}{\textbf{AUROC}} & \multicolumn{1}{c}{\textbf{F1}} \\

\midrule

AASIST       & 1.00\%                           & 0.9992                                & 0.9755                          \\
RawGAT-ST    & 0.60\%                           & 0.9984                                & 0.9869                          \\
RawNet2      & 5.20\%                           & 0.9876                                & 0.9505                          \\
Res-TSSDNet  & 1.20\%                           & 0.9995                                & 0.9900                          \\
SAMO         & 5.20\%                           & 0.9780                                & 0.9041                          \\
AST-asvspoof & 6.20\%                           & 0.9814                                & 0.9370                         \\
\bottomrule
\end{tabular}
\end{table*}

\paragraph{In-Distribution Detection on ASVspoof 2019.}
Benchmarking six dedicated detectors on a 1k subset of ASVspoof 2019 reveals that all methods achieve AUROC scores exceeding 0.978, indicating the benchmark is largely saturated under controlled conditions. However, EER and F1 scores highlight meaningful performance gaps. As shown in Table~\ref{tab:asvspoof_results}, Res-TSSDNet achieves the best overall balance (1.20\% EER, 0.9995 AUROC, 0.9900 F1), while RawGAT-ST attains the lowest strict EER at 0.60\%. AASIST provides solid competitive results with a 1.00\% EER. Conversely, SAMO, AST-asvspoof, and RawNet2 exhibit elevated EERs between 5.20\% and 6.20\%. Despite their respectable AUROC values, these high EERs indicate poor threshold calibration, underscoring that AUROC alone is an overly optimistic metric for acoustic detection.

\begin{table*}[h]
\centering
\small

\caption{EER (\%), AUC-ROC, and F1 score of AntiDeepfake variants (Wav2Vec2-Large, HuBERT-XLarge, XLS-R 2B) and XLS-R + SLS evaluated across three datasets of increasing ecological validity: ASVspoof 2019, FoR, and In-the-Wild. Results reproduce the experimental scenario of Ge et al.\cite{ge2025antideepfake}. Lower EER indicates better performance; higher AUC-ROC and F1 indicate better performance.}
\label{tab:antideepfake_results}

\resizebox{\textwidth}{!}{%
\begin{tabular}{clccclccclccc}
\toprule
\textbf{Dataset $\rightarrow$}                     &  & \multicolumn{3}{c}{\textbf{ASVspoof 2019 (1k)}}                                                            &  & \multicolumn{3}{c}{\textbf{FoR (1k)}}                                                                      &  & \multicolumn{3}{c}{\textbf{In-the-Wild (1k)}}                                                              \\

\cmidrule(lr){1-1}
\cmidrule(lr){3-5}
\cmidrule(lr){7-9}
\cmidrule(lr){11-13}

\textbf{Detector $\downarrow$ / Metric $\rightarrow$}         &    & \multicolumn{1}{c}{\textbf{EER}} & \multicolumn{1}{c}{\textbf{AUROC}} & \multicolumn{1}{c}{\textbf{F1}} &  & \multicolumn{1}{c}{\textbf{EER}} & \multicolumn{1}{c}{\textbf{AUROC}} & \multicolumn{1}{c}{\textbf{F1}} &  & \multicolumn{1}{c}{\textbf{EER}} & \multicolumn{1}{c}{\textbf{AUROC}} & \multicolumn{1}{c}{\textbf{F1}} \\

\midrule

AD(Wav2Vec2-Large) &  & 0.20\%                           & 0.9970                                & 0.9970                          &  & 7.20\%                           & 0.9738                                & 0.8461                          &  & 1.60\%                           & 0.9984                                & 0.9832                          \\
AD(HuBERT-XLarge)  &  & 0.00\%                           & 1.0000                                & 0.9990                          &  & 11.80\%                          & 0.9554                                & 0.8084                          &  & 4.80\%                           & 0.9934                                & 0.9326                          \\
AD(XLS-R 2B)       &  & 0.60\%                           & 0.9999                                & 0.9970                          &  & 8.80\%                           & 0.9774                                & 0.8399                          &  & 1.20\%                           & 0.9987                                & 0.9860                          \\
XLS-R + SLS                   &  & 0.40\%                           & 0.9995                                & 0.9442                          &  & 11.80\%                          & 0.9567                                & 0.7868                          &  & 12.80\%                          & 0.9413                                & 0.7558                            \\
\bottomrule
\end{tabular}
}
\end{table*}

\paragraph{Cross-Dataset Generalization of Foundation Models.}
We subsequently evaluate the out-of-distribution robustness of large pretrained speech models. All four foundation models perform exceptionally well in-distribution on ASVspoof 2019 as reported in Table~\ref{tab:antideepfake_results}, with HuBERT-XLarge achieving a perfect 0.00\% EER. However, performance degrades substantially across all models on the FoR dataset. Wav2Vec2-Large exhibits the most robust transfer (7.20\% EER), whereas HuBERT-XLarge suffers a catastrophic drop to an 11.80\% EER. This suggests severe overfitting to ASVspoof-specific artifacts despite its massive model capacity.

On the highly naturalistic In-the-Wild dataset, XLS-R-2B achieves the best overall results (1.20\% EER, 0.9987 AUROC), proving to be the most consistent and robust model across all tested domains. Notably, the XLS-R + SLS baseline collapses entirely on this dataset to an EER of 12.80\% and an F1 of 0.7558. This failure highlights that specific supervised training objectives, while highly effective on controlled benchmarks, do not necessarily transfer to real-world, naturalistic audio deepfakes.

% \subsection{Per-Detector Computational Requirements}
% \label{app:compute}

% Computational costs vary significantly across detector families:
% \begin{itemize}[leftmargin=1.5em, itemsep=1pt]
%     \item \textbf{Statistical methods} (log-likelihood, entropy, rank variants): Single forward pass through a causal LM. Fastest category ($\sim$50--200 samples/sec on A100 with GPT-2).
%     \item \textbf{Perturbation-based} (DetectGPT): Require multiple forward passes per sample (one per perturbation). 5--20$\times$ slower than statistical methods.
%     \item \textbf{Fast perturbation-free} (Fast-DetectGPT, Glimpse): Single forward pass with distribution sampling. Comparable to statistical methods.
%     \item \textbf{Multi-model} (Binoculars): Two model forward passes per sample. 2$\times$ the memory of single-model methods.
%     \item \textbf{Supervised} (RoBERTa, RADAR): Single forward pass through a fine-tuned classifier. Fast inference ($\sim$200--500 samples/sec).
%     \item \textbf{Image detectors}: Vary from lightweight (CNNSpot: ResNet-50 inference) to heavy (AEROBLADE: VAE encode-decode + LPIPS computation).
%     \item \textbf{Audio detectors}: Generally lightweight once the model is loaded ($\sim$100--300 clips/sec for RawNet2, AASIST).
% \end{itemize}

% =====================================================================
% NeurIPS Checklist
% =====================================================================
% \clearpage

% \input{checklist.tex}
 
\end{document}